\documentclass[%
 reprint,
 superscriptaddress,
 nofootinbib,
 amsmath,amssymb,
 aps,
 prx,
 floatfix,
]{revtex4-2}
\usepackage{graphicx}%
\usepackage{bm}
\usepackage{braket}
\usepackage[hidelinks,colorlinks=true]{hyperref}
\usepackage{stmaryrd}%
\usepackage{multirow, makecell}
\usepackage[ruled,lined]{algorithm2e}

\usepackage{tikz}
\usetikzlibrary{quantikz2}%
\newcommand{\phiest}{\tilde{\phi}}
\newcommand{\im}{\mathrm{i}}
\newcommand{\e}{\mathrm{e}}
\newcommand{\bE}{\mathbb{E}}

\newcommand{\Ns}{N_{\rm s}}
\newcommand{\direct}{\overline{R}^{\rm D}}
\newcommand{\rgt}{\overline{R}^{\rm RGT}}
\newcommand{\rgtQED}{\overline{R}^{\rm RGT}}

\begin{document}

\title{Quantum Error-Corrected Computation of Molecular Energies}

\def\quantinuumTokyo{Quantinuum K.K., Otemachi Financial City Grand Cube 3F, 1-9-2 Otemachi, Chiyoda-ku, Tokyo, Japan}
\def\quantinuumColorado{Quantinuum, LLC, 303 South Technology Court, Broomfield, Colorado 80021, USA}
\def\riken{RIKEN Center for Interdisciplinary Theoretical and Mathematical Sciences (iTHEMS), RIKEN, Wako, Saitama 351-0198, Japan}
\def\quantinuumCambridge{Quantinuum Ltd., Terrington House, 13-15 Hills Road, Cambridge CB2 1NL, UK}

\author{Kentaro~Yamamoto}
\email{kentaro.yamamoto@quantinuum.com}
\affiliation{\quantinuumTokyo}

\author{Yuta~Kikuchi}
\affiliation{\quantinuumTokyo}
\affiliation{\riken}

\author{David~Amaro}
\affiliation{\quantinuumCambridge}

\author{Ben~Criger}
\affiliation{\quantinuumCambridge}

\author{Silas~Dilkes}
\affiliation{\quantinuumCambridge}

\author{Ciar\'{a}n~Ryan-Anderson}
\affiliation{\quantinuumColorado}

\author{Andrew~Tranter}
\affiliation{\quantinuumCambridge}

\author{Joan~M.~Dreiling}
\affiliation{\quantinuumColorado}

\author{Dan~Gresh}
\affiliation{\quantinuumColorado}

\author{Cameron~Foltz}
\affiliation{\quantinuumColorado}

\author{Michael~Mills}
\affiliation{\quantinuumColorado}

\author{Steven~A.~Moses}
\thanks{Current Affiliation: AWS Center for Quantum Computing, Pasadena CA 91125}
\affiliation{\quantinuumColorado}

\author{Peter~E.~Siegfried}
\affiliation{\quantinuumColorado}

\author{Maxwell~D.~Urmey}
\affiliation{\quantinuumColorado}

\author{Justin~J.~Burau}
\affiliation{\quantinuumColorado}

\author{Aaron~Hankin}
\affiliation{\quantinuumColorado}

\author{Dominic~Lucchetti}
\affiliation{\quantinuumColorado}

\author{John~P.~Gaebler}
\affiliation{\quantinuumColorado}

\author{Natalie~C.~Brown}
\affiliation{\quantinuumColorado}

\author{Brian~Neyenhuis}
\affiliation{\quantinuumColorado}

\author{David~Mu\~noz Ramo}
\affiliation{\quantinuumCambridge}

\date{}

\begin{abstract}

We present the first demonstration of an end-to-end pipeline with quantum error correction (QEC) for a quantum computation of the electronic structure of molecular systems. We calculate the ground-state energy of molecular hydrogen, using quantum phase estimation (QPE) on qubits encoded with the $\llbracket7,1,3\rrbracket$ color code on Quantinuum H2--2.
We obtain improvements in computational fidelity by
(1) introducing several partially fault-tolerant (FT) techniques for the Clifford+$R_{Z}$ (arbitrary-angle single-qubit rotation) gate set, and
(2) integrating Steane QEC gadgets for real-time error correction.
In particular, the latter enhances the QPE circuits' performance despite the complexity of the extra QEC circuitry.
The encoded circuits contain up to 1585 (546) fixed and 7202 (1702) conditional physical two-qubit gates (mid-circuit measurements), and $\sim$3900 ($\sim$760) total operations are applied on average.
The energy $E$ is experimentally estimated to within $E - E_{\mathrm{FCI}} = 0.001(13)$ hartree,
where $E_{\mathrm{FCI}}$ denotes the exact ground state energy within the given basis set.
Additionally, we conduct numerical simulations with tunable noise parameters to identify the dominant sources of noise. We find that orienting the QEC protocols towards higher memory noise protection is the most promising avenue to improve our experimental results.
\end{abstract}

\maketitle

\section{Introduction\label{sec:introduction}}

Computational chemistry is a particularly promising application of quantum computers, due to the known exponential quantum speedup, with a modest scaling pre-factor in terms of the molecular system size~\cite{Aspuru-Guzik2005-tv,OMalley2016-py,Cao2019-tn,McArdle2020-rs,Chan2024-km}. Quantum phase estimation (QPE)~\cite{Nielsen2010-ki} is a critical subroutine to achieve this speedup. The basic QPE workflow infers the eigenvalue $\e^{\im\phi_{j}}$ of a given unitary $U$ associated with the eigenstate $\ket{\phi_{j}}$~\cite{Nielsen2010-ki}. In the case of a quantum chemistry simulation, $U$ is the time evolution operator $\e^{-\im Ht}$ for a particular time $t$, where $H$ is the system Hamiltonian. This time evolution is coherently controlled using a uniform superposition state on a register of ancilla qubits, which produces a number of interfering time series that can be processed with an inverse quantum Fourier transform.

The textbook version of QPE~\cite{Nielsen2010-ki} uses a very deep circuit that rapidly accumulates noise during runtime, rendering implementing the algorithm significantly challenging except for the most trivial cases. Several proposals have been presented lately to reduce resource requirements. They range from techniques to reduce the complexity of the time evolution circuit~\cite{Berry2019-mj,Campbell2019-fc,Nakaji2024-yt,Koczor2024-sx,Lee2021, caesura_arxiv2025} to implementations where the ancilla register is reduced to just one qubit at the cost of increasing the number of samples required from the quantum computer and addition of extra classical postprocessing steps~\cite{kitaev1995quantum, kitaev2002classical, Knill2007, Higgins2007, Svore2014-ka, Wiebe2016, OBrien2019-ic, Somma2019, Lin2022, Wan2022, Ding2023-cb,Yamamoto2024-vm, Paul2025-gz}. Recent work even proposes a version of QPE without an ancilla register~\cite{Clinton2024-sc}.
It is fair to say that these proposals, while helpful, still yield QPE quantum circuits that require unrealistic physical error rates for noisy quantum hardware. For this reason, a scalable implementation of this algorithm requires the use of quantum error correction (QEC).

In QEC, logical qubits are encoded in a protected subspace of a higher-dimensional Hilbert space, typically that of a larger number of physical qubits. This redundancy allows for the detection and correction of errors without directly measuring or disturbing the logical information. The quantum gates operating in the circuit are also encoded at the logical level, and error syndromes are measured periodically, enabling the identification of errors and the application of corrections~\cite{Gottesman1997-ka}. Replacing physical qubits with logical qubits ensures that multiple independent errors on different qubits are necessary in order to corrupt the logical information, resulting in a smaller overall error probability.

Though QEC is necessary for implementing scalable, complex algorithms like QPE, there is a tradeoff between the error suppression of noise and the resources required to implement a QEC protocol.
In addition to an increase in the number of physical qubits required, the circuits have to be designed in a fault-tolerant (FT) manner to keep the spread of errors under control.
There are implementations of QEC protocols and gates that are naturally FT (i.e., transversal gates, Steane QEC gadgets~\cite{Steane1996}), but more often than not, one must tailor protocols to handle error propagation by catching errors before they can spread.
This often comes at the cost of increasing the circuit overhead in terms of physical circuit depth and number of physical qubits.

In practice, however, the large overhead often makes FT protocols underperform other less demanding designs. Partial FT designs aim to significantly reduce the largest sources of overhead at the cost of sacrificing strict fault-tolerance~\cite{Dangwal2025, Toshio2024, Akahoshi2024, Self2022_natphys}. For example, in~\cite{Toshio2024, Akahoshi2024} the authors replace the resource-intensive FT magic state distillation and gate synthesis protocols with a much lighter non-FT preparation and injection of arbitrary-angle rotations in the surface code. Similarly, identifying hardware features, such as native gate sets and noise characteristics, can allow for tailored QEC protocols that minimize error propagation while avoiding unnecessary resource overhead.

There are examples in the literature of implementations for different components of QEC protocols, including logical state preparation, memory, and gates~\cite{Mayer2024-el,Reichardt2024-wa,Da_Silva2024-da,Reichardt2024-jt,Miao2023-kn,Google-Quantum-AI-and-Collaborators2025-mv,Bluvstein2023-tp,Caune2024-jt,Lacroix2024}. However, a full experimental implementation of a QEC-encoded algorithm for chemistry applications is still lacking, especially for such an important method as QPE.
{Previous logical-level experiments have focused on demonstrating fault-tolerant components, such as logical quantum memory (e.g., \cite{Google-Quantum-AI-and-Collaborators2025-mv}), logical state preparation, and high-fidelity logical gates. Other demonstrations have implemented logical subroutines, such as the logical QFT~\cite{Mayer2024-el}, which are valuable building blocks but are not themselves full algorithms.}
This paper reports progress in addressing this gap {by integrating both levels—combining these fault-tolerant components and subroutine-level capabilities—into a complete logical algorithm designed to perform a meaningful quantum-chemistry calculation}~\cite{Blunt2024-bj}. We demonstrate an end-to-end QEC workflow using a Quantinuum trapped-ion quantum computer, System Model H2~\cite{Pino2021-ck,Moses2023-qw}, showcasing the application of QPE with the $\llbracket7,1,3\rrbracket$ color code for accurate ground-state energy calculations.
Our approach integrates advanced circuit compilation techniques, such as recursive gate teleportation~\cite{Trout2015-nr}, to achieve accurate logical quantum computations despite the limitations of current quantum hardware. We use the Hamiltonian corresponding to the hydrogen molecule with a minimal basis set~\cite{Szabo1996-rg,Helgaker2014-ce} at equilibrium geometry to experimentally demonstrate this workflow and calculate its ground-state energy using a single-ancilla version of QPE to reduce resources needed for the computation.

Beyond experimental validation, our end-to-end QEC pipeline facilitates numerical exploration of algorithmic performance with the noise model appropriately arranged for future hardware.
To better understand the impact of hardware noise on a partially fault-tolerant QPE circuit, we conduct numerical simulations with tunable noise parameters, enabling systematic evaluation of the trade-offs between resource overhead and circuit fidelity.
While two-qubit gate infidelity is commonly considered a dominant error source, our results suggest that errors arising during extended idling and ion transport—often referred to as memory noise—can play a more significant role in encoded circuits with substantial additional circuitry.
This exploration highlights opportunities to improve experimental results via QEC and error mitigation protocols, as well as hardware development oriented toward addressing memory noise.

The rest of the paper is organized as follows.
In Sec.~\ref{sec:method}, we describe the main components of our workflow. This includes a description of the version of QPE, the $\llbracket7,1,3\rrbracket$ color code, compilation of the quantum circuit according to the native gate set admitted by the code and calibration procedure.
Section~\ref{sec:experiment} deals with the setup chosen for the experimental realization of the workflow and the results obtained from running it on our trapped-ion device.
Section~\ref{sec:numerical} explains the procedure to investigate the main contributors to the performance of the circuit execution.
Finally,  Sec.~\ref{sec:conclusions} provides a summary of our results and an outlook for the efficient implementation of this workflow.

\section{Method\label{sec:method}}

\begin{figure}[!htbp]
    \centering
    \includegraphics[width=0.99\hsize]{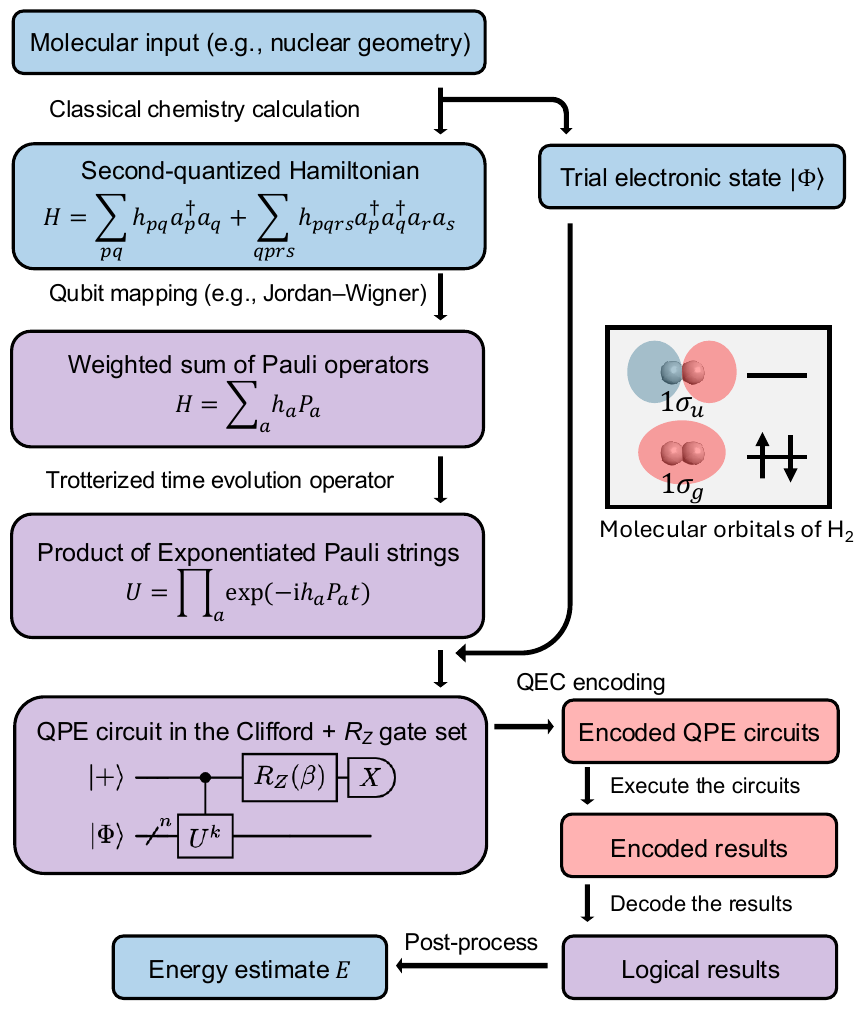}
    \caption{
        Overview of the end-to-end quantum computational chemistry workflow.
        Three different types of data are involved, namely, chemistry (blue), quantum information at a logical level (purple), and that at an encoded physical level (red).
        Molecular orbitals of the target system, molecular hydrogen H$_{2}$, are shown to facilitate the chemical interpretation.
    }
    \label{fig:workflow}
\end{figure}

An overview of the present end-to-end workflow is shown in Fig.~\ref{fig:workflow}.
We note that similar workflows have recently been studied in~\cite{Trout2015-nr,Blunt2024-bj} and demonstrated in~\cite{Dam2024-ed}, where variational algorithms are used to calculate energies~\cite{Cerezo2021-lc} with the circuits protected by quantum error detection. Our approach is differentiated from these through the application of QPE, QEC, and partial fault tolerance.

The workflow starts with classical computational chemistry calculations to construct a molecular Hamiltonian $H$ and a trial electronic state, such as a Hartree--Fock state~\cite{Helgaker2014-ce, Szabo1996-rg}.
We consider the single-qubit Hamiltonian
\begin{align}
    H = h_{1} Z +h_{2} X +h_{3} I
    ,
    \label{eq:hamiltonian}
\end{align}
with $(h_{1}, h_{2}, h_{3}) = (0.79605, -0.18092, -0.32096)$ in atomic units ~\cite{Helgaker2014-ce}.
These parameters are evaluated with \texttt{InQuanto} v4.2~\cite{inquanto} and its extension to \texttt{PySCF} v2.4~\cite{Sun2018-gy} by performing the restricted Hartree--Fock method~\cite{Helgaker2014-ce,Szabo1996-rg}, {Jordan--Wigner transformation~\cite{McArdle2020-rs}, and tapering off three qubits exploiting symmetry~\cite{Bravyi2017-xh}.}
The Hamiltonian~\eqref{eq:hamiltonian} describes the minimal-basis hydrogen molecule at the equilibrium geometry.
The circuit representing controlled-$U$ and the preparation of $\ket{\Phi}$ are then constructed using $H$ and the trial electronic state, respectively~\cite{Cao2019-tn,McArdle2020-rs}.
Next, we construct the QPE circuits~\cite{Svore2014-ka} with these components (Sec.~\ref{subsec:qpe}) and write them in terms of logical gates for the $\llbracket7,1,3\rrbracket$ color code~\cite{Nielsen2010-ki,pecos} (Sec.~\ref{ssec:steane}).
Finally, we run the encoded circuits and post-process the measurement outcomes to obtain the estimated energy.
These steps are detailed in the following subsections.

\subsection{Quantum phase estimation\label{subsec:qpe}}

Given a quantum state
\begin{align}
\label{eq:trial_state}
    \ket{\Phi}
    =
    \sum_{j=0}^{N_{\rm eig}-1}a_{j}\ket{\phi_{j}}
\end{align}
and the unitary $U=\e^{-\im Ht}$, the task is to estimate a subset of the eigenphases $\{\phi_j\}_{j=0}^{N_{\rm eig}-1}$. The eigenstates are labeled in the population descending order, i.e., $|a_j|^2 \ge |a_{j+1}|^2$.
When the input state $\ket{\Phi}$ approximates the ground state, $-\phi_0/t$ is most likely the ground state energy.
In the present work, we use information theory QPE~\cite{Svore2014-ka} to estimate the phase $\phi_0$.

The QPE uses the following circuit:
\begin{align}
\label{eq:qpe_circ}
\begin{quantikz}[style={column sep=4pt, row sep=5pt}]
    \lstick{$\ket{+}$}
    &
    &
    &
    & \ctrl{1}
    & \gate{R_{Z}(\beta)}
    & \meterD{X}
    & \cw\rstick{$m$}
    \\
    \lstick{$\ket{\Phi}$}
    &
    & \qwbundle{n}
    &
    & \gate{U^{k}}
    &
    &
\end{quantikz}
\end{align}
This circuit has four key components:
\begin{enumerate}
\item Preparation of $\ket{\Phi}$ and $\ket{+}$.
\item Application of $U^k$ on the system register conditioned on the ancilla qubit (ctrl-$U^k$).
\item Application of a phase-shift gate, {$R_Z(\beta):=\e^{-\im\frac{\beta}{2}Z}$.}
\item Measurement of the ancilla qubit in the $X$-basis.
\end{enumerate}
The parameters $\{k_{l}\}$ and $\{\beta_{l}\}$ are randomly sampled from $k \in \{1,2,\dots, k_{\mathrm{max}}\}$ and $\beta \in \{0, \frac{\pi}{2}\}$, respectively, where $k_{\mathrm{max}}$ controls the precision of QPE, and determines the maximum depth of the circuit~\eqref{eq:qpe_circ}.
The probability of measuring $m \in \{0, 1\}$ from a single ideal execution of the circuit~\eqref{eq:qpe_circ} is expressed as,
\begin{align}
\label{eq:prob_circ}
    P(m|\{a_j, \phi_j\}; k,\beta)
    =
    \sum_{j=0}^{N_{\rm eig}}
    |a_j|^2
    \frac{1 + \cos(k\phi_{j} + \beta - m\pi)}{2}.
\end{align}
We run the circuit~\eqref{eq:qpe_circ} $N_{\mathrm{s}}$ times and collect the measurement outcomes $\{m_{l}\}_{l=1}^{\Ns}$.

Provided $\{m_{l}\}_{l=1}^{\Ns}$, we calculate the following distribution over $\phiest\in[0,2\pi)$,
\begin{align}
\label{eq:probs}
    &Q(\phiest\mid \{m_{l}\}; \{k_{l},\beta_{l}\})
    :=
    \prod_{l=1}^{N_{\mathrm{s}}}
    Q(m_{l}\mid \phiest; k_{l},\beta_{l})
    \\
\label{eq:probs_single}
    &\text{ with} \quad
    Q(m_l\mid\phiest; k_l,\beta_l)
    =
    \frac{1 + \cos(k_l\phiest + \beta_l - m_l\pi)}{2}.
\end{align}
The expression~\eqref{eq:probs_single} would represent the probability of obtaining a measurement outcome $m_l$ if the input state were an eigenvector with the eigenphase $\phiest$.
For an approximate ground state $\ket{\Phi}$~\eqref{eq:trial_state}, the distribution~\eqref{eq:probs} is peaked around $\{\phi_j\}$ with the most dominant peak at $\phi_0$.
Hence, one can estimate $\phi_0$ by the phase $\phiest$ that maximizes~\eqref{eq:probs},
\begin{align}
\label{eq:argmax}
    \underset{\phiest}{\rm arg\,max\,} Q(\phiest|\{m_{l}\}; \{k_{l},\beta_{l}\}).
\end{align}
The accuracy of estimate depends on the parameters $k_{\rm max}$ and $\Ns$, and the population $|a_0|^2$ of $\ket{\phi_0}$ in the input state~\eqref{eq:trial_state}. See App.~\ref{app:ipqe} for details.

\vspace{1em}

For the ground-state energy estimation of the Hamiltonian~\eqref{eq:hamiltonian}, we approximate the time evolution operator $\e^{-\im Hkt}$ up to the global phase $\e^{-\im h_{3}kt}$ by the Lie-Trotter formula~\cite{Lloyd1996-mw}
\begin{eqnarray}
\label{eq:Trotter}
    U^{k}
    &:=&
    (\e^{-\im h_{1}Zt}\e^{-\im h_{2}Xt})^{k}
    \\
    \nonumber
    &=&
    \e^{-\im (H-h_{3}I)kt} + {\cal O}(kt^{2})
    .
\end{eqnarray}
In this work, the evolution time is set to $t=\pi/(8h_1)$, leading to $\e^{-\im h_{1}Zt} = \e^{-\im\frac{\pi}{8}Z} = T$.
Moreover, we approximate $h_{2}t/\pi$ with a 5-bit binary fraction, so that its partially FT encoding is simplified (see Sec.~\ref{sec:encoding}). The truncation results in a systematic error of $\sim$$10^{-3}$ hartree.

The input state $\ket{\Phi}$ is parameterized by
\begin{align}
    \ket{\Phi(\alpha_{0}, \alpha_{1})}
    =
    R_{Z}(\alpha_{1}) R_{Y}(\alpha_{0})\ket{1}
    \label{eq:ansatz},
\end{align}
with $\alpha_{0}, \alpha_{1} \in [0, 2\pi)$.
They are set to $(\alpha_{0}, \alpha_{1}) = (-0.274220, -0.785398)$ for the calibration conducted in Sec.~\ref{ssec:calibration}, at which the state~\eqref{eq:ansatz} approximates the ground state of $H$ up to the Trotter error.
For the ground-state energy estimation by QPE, we set the parameters to $(\alpha_{0}, \alpha_{1}) = (0, 0)$ so that the state~\eqref{eq:ansatz} represents the Hartree--Fock state.
{We note that preparing an approximate ground state becomes more costly as the geometry deviates from equilibrium. Nevertheless, we only consider the Hartree--Fock input state near the equilibrium geometry for the purpose of benchmarking the QPE.}

\subsection{Partial fault-tolerance and logical arbitrary angles}
\label{ssec:steane}

\begin{figure}[!htbp]
    \centering
    \includegraphics[width=0.75\linewidth]{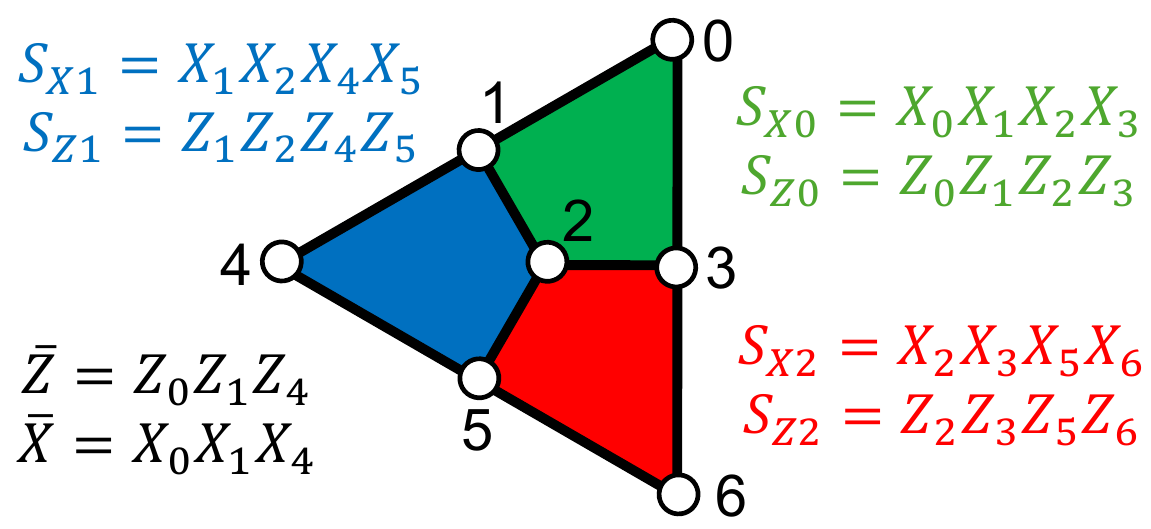}
    \caption{
        Convention of the stabilizer generators and logical operators for the $\llbracket 7, 1, 3\rrbracket$ color code in the present work. Physical qubits are indicated by white circles. Logical operators and stabilizer generators are defined as tensor products of single-qubit Pauli operators.
        For each cell, the four qubits at the vertices define both $X$-type and $Z$-type stabilizer generators.
    }
    \label{fig:steane}
\end{figure}

The $\llbracket 7, 1, 3\rrbracket$ color code~\cite{Steane1997-tc,Ryan-Anderson2021-me} takes 7 physical qubits to encode 1 logical qubit to distance 3, enabling the correction of any single-qubit errors. In what follows, we refer to the $\llbracket7,1,3\rrbracket$ color code as the color code.
The generating set of stabilizers and logical operators are given in Fig.~\ref{fig:steane} with the overhead bar indicating the logical operation.

The QPE circuit encoded with the color code consists of various components: logical state preparation, logical gates, QEC cycles, and final measurement. The logical error rate of the QPE circuit increases with the logical error rate of its components, so care must be taken in their design.
In what follows, we motivate the use of partial fault-tolerant circuits for logical gates in the color code and present four implementations for arbitrary logical rotations, each with a different level of FT.

\subsubsection{Partial-fault tolerance}

FT designs assume that each physical operation (e.g., gates, measurements) can fail with a small probability $p$, introducing random Pauli errors on affected qubits, but such a failure does not propagate to an undetectable (and thus uncorrectable) logical error.
For distance three, this means that circuits are designed so that every single fault results in a non-problematic error: a detectable error or a trivial error of weight-1 up to the stabilizers of the code.
This ensures that at least two faults are necessary to produce a logical error: they can accumulate into an uncorrectable weight-2 error.
Under the FT design, the probability of a logical error scales as $Ap^2+O(p^3)$, where $A$ is the number of pairs of faults that produce a logical error.
This is in contrast to non-FT components, where the logical error rate scales as $Bp + O(p^2)$, where $B$ is the number of single faults that produce a logical error.

Transversal gates are naturally FT, as faults do not spread to more qubits than they originally affect.
For example, the logical Hadamard in the color code is implemented as a tensor product of 7 physical Hadamards,
$\overline{H}= H^{\otimes 7}$, ensuring that any single fault results in at most a weight-1 Pauli error.
In the color code, all logical Clifford operations admit a transversal implementation, including the logical phase gate, $\overline{S}= (S^{\dagger})^{\otimes 7}$, and the logical CNOT, $\overline{CX}$~\cite{Nielsen2010-ki}.

Compared to a FT implementation, a non-FT approach implements the operation without accounting for how errors occur or propagate, potentially leading to uncorrectable logical errors.
A trivially non-FT component is the direct implementation of a logical gate as a three-body rotation: $\direct_Z(\theta)$ presented in Tab.~\ref{tab:RGT_rotation}.
It consists of three physical operations: two $CX$ gates and a $R_{ZZ}(\theta)=\e^{-\im\frac{\theta}{2}ZZ}$ rotation.
For instance, the failure of the second $CX$ gate can introduce a weight-2 error. A more comprehensive error analysis is found in App.~\ref{app:Z_rotation}.

Finding efficient FT components, such as transversal gates, can be challenging.
Often in practice, non-FT components are made FT by adding extra overhead (i.e., flag qubits, detection gadgets) to catch malignant faults before they can spread.
Consequently, for small code distances like $d=3$, despite their protection against single faults, FT components might be less protected against two or more faults than non-FT components: the prefactor $A$ grows faster with the overhead than $B$.
In practice, in the regime of $p$ not too small and FT gadgetry overhead large, the logical error rate of the FT component might underperform the logical error rate of the non-FT component: $Ap^2 > Bp$.
In this context, relaxing the strict FT design to tolerate logical errors caused by some single faults might be beneficial to reduce the logical error rate if it allows a significant reduction of the overhead~\cite{Trout2015-nr}.

Such a design principle, i.e., the addition of minimal overhead that reduces the logical error rate without producing strictly FT components, has been referred to as \textit{partial fault-tolerance}.
Partial FT components have been proposed in~\cite{Akahoshi2024, Toshio2024, Dangwal2025} and successfully demonstrated~\cite{Self2022_natphys, Yamamoto2024-vm}.
In this work, we compile the QPE circuit into the Clifford+$\overline{R}_Z(\theta)$ gateset~\cite{Akahoshi2024, Toshio2024} and propose partially FT logical rotations of an arbitrary angle $\overline{R}_Z^{\mathrm{RGT}}(\theta)$ for the Steane code presented in Table~\ref{tab:RGT_rotation}. This arbitrary-angle logical rotation is the component that benefits the most from a partial FT design, as its FT implementation requires a complicated process of magic state distillation~\cite{Gidney2024, Gidney2019, Litinski2019, Bravyi2005} and gate synthesis~\cite{Ghosh2025, Kliuchnikov2013, Kitaev1997} into the more limited Clifford+$T$ gate set that significantly increases the overhead.

\subsubsection{Arbitrary-angle logical rotations}

\begin{table*}[htb]
    \centering
    \def\arraystretch{1.5}
    \begin{tabular}{c||c|c|c|c}
        \hline
        & \qquad RGT \enspace\enspace
        & $\ket{\overline{\theta}}$ preparation
        & Measurement
        & Circuit
        \\
        \hline\hline
        \rule{0pt}{3.5em}
        $\direct_{Z}(\theta)$
        & No
        & N/A
        & N/A
        &
        $\begin{quantikz}[style={column sep=2pt, row sep=2pt}]
            \lstick{} & \gate[style={fill=blue!10}]{\direct_{Z}(\theta)} & \qw
        \end{quantikz}
        ~=~
        \begin{quantikz}[style={column sep=2pt, row sep=2pt}]
             & \qw       & \gate[2]{R_{ZZ}(\theta)} & \qw       & \qw \\
             & \targ{}   &                         & \targ{}   & \qw \\
             & \ctrl{-1} & \qw                     & \ctrl{-1} & \qw \\
        \end{quantikz}$
        \\
        \hline
        \rule{0pt}{3.5em}
        $\rgtQED_Z(\theta)$
        & Yes
        & non FT + QED
        & FT
        &$\begin{quantikz}[style={column sep=4pt, row sep=5pt}]
            \lstick{} & \gate[style={fill=green!15}]{\rgtQED_Z(\theta)} & \qw
        \end{quantikz}
        ~=~
        \begin{quantikz}[style={column sep=4pt, row sep=5pt}]
            \lstick{}                          & \ctrl{1}            & \gate[style={fill=green!15}]{\rgtQED_Z(2\theta)} & \qw \\
            \lstick{$\ket{\overline{\theta}^{\rm QED}}$} & \gate{\overline{X}} & \meterD{\overline{Z}} \vcw{-1} \\
        \end{quantikz}$
        \\
        \hline
    \end{tabular}
    \caption{\label{tab:RGT_rotation}
    Implementations of a logical rotation $\overline{R}_Z(\theta)$. The top implementation does not use the recursive gate teleportation (RGT), and thus does not need $\ket{\overline{\theta}}$ preparation and measurement. The state $\ket{\overline{\theta}}$ in the second row is encoded non-fault-tolerantly with \eqref{eq:partialFT_resource_prep}, denoted by $\ket{\overline{\theta}^{\rm QED}}$. The bottom register is measured in $\overline{Z}$-basis with the component~\eqref{eq:circ_meas}. In the RGT circuit, a controlled $\overline{X}$ gate diagram stands for a transversal $CX$.
    }
\end{table*}

We now present our implementations of an arbitrary-angle logical rotation as summarized in Table~\ref{tab:RGT_rotation}. They are represented by colored boxes in quantum circuit diagrams.

The direct implementation $\direct_Z(\theta)$ adds the least amount of overhead with only three two-qubit gates.
One can add a low-overhead and specialized QED component to $\direct_Z(\theta)$ to detect some of the problematic errors in it (see Sec.~\ref{app:Z_rotation} for error analysis). Depicted in circuit~\eqref{eq:circ_QED}, this specialized QED component measures the set of stabilizers $\{S_{X0}, S_{X1}, S_{Z0}, S_{Z1}\}$ in Fig~\ref{fig:steane} and discards the entire circuit on the detection of an error.
This QED component avoids the need for flag qubits by carefully ordering the CNOTs so that the ancilla qubits act as the flag of each other~~\cite{Chao2018,Self2022_natphys}.

The other implementation, $\rgtQED_Z(\theta)$, consists of the non-FT preparation of non-stabilizer states $\ket{\overline{\theta}} = \e^{-\im \frac{\theta}{2} \overline{Z}}\ket{\overline{+}}$ and their recursive gate teleportation (RGT)~\cite{Trout2015-nr, Mayer2024-el, Toshio2024, Akahoshi2024} into the computational qubit:
\begin{align}
\label{eq:circ_RGT}
    \begin{quantikz}[style={column sep=4pt, row sep=7pt}]
        \lstick{$\ket{\psi}$}
        &
        & \ctrl{1}
        &
        & \gate{R_{Z}(2\theta)}
        &
        & \rstick{$R_{Z}(\theta)\ket{\psi}$}
        \\
        \lstick{$\ket{\theta}$}
        &
        & \targ{}
        &
        & \meterD{Z} \vcw{-1}
    \end{quantikz}
\end{align}
Gate teleportation is understood by considering the state before the measurement,
\begin{align}
    \frac{1}{\sqrt{2}}R_{Z}(\theta)\ket{\psi}\otimes \ket{0}
    +
    \frac{1}{\sqrt{2}}R_{Z}(-\theta)\ket{\psi}\otimes \ket{1}.
\end{align}
Measuring ``0'' in the second register (bottom register in circuit~\eqref{eq:circ_RGT}) results in the desired state, and otherwise, the rotation angle is applied with an opposite sign. Each case happens with probability $1/2$. If the latter happens, this protocol with the rotation angle $2\theta$ is applied to correct the angle to $\theta$. This procedure is repeated until ``0'' is measured to indicate a successful application of $R_{Z}(\theta)$. Provided a rotation angle represented by an $n_b$-bit binary fraction, i.e. $\theta/\pi=b_0+b_1/2+\dots b_{n_b}/2^{n_b-1}$ with $b_i\in\{0,1\}$, the protocol succeeds after at most $n_b-2$ recursions. This is because $R_Z(2^{n_b-2}\theta)=R_Z(\pi(b_{n_b-1}+ b_{n_b}/2))$ is a Clifford gate, which can be applied transversely, and thus does not require gate teleportation.

The state preparation $\ket{\overline{\theta}^\mathrm{QED}}$ consists of the FT preparation of $\ket{\overline{0}}$ from~\cite{Goto2016-rl} followed by a logical Hadamard, the non-FT $\direct_Z(\theta)$, and a QED round that discards on the detection of any error. This process is repeated until no error is detected. If the repetition reaches $r_{\rm rus}$ times without success, the circuit execution is aborted, leading to a sample loss. We choose $r_{\rm rus}=2$ in this work.

{We remark on the advantage of the RGT scheme over the direct implementation. Faults in the direct implementation of $\overline{R}^{\rm D}_Z(\theta)$ lead to various error propagations (see Table~\ref{tab:errors}). All the weight-2 and higher weight errors except one are detected by a trailing QED gadget (making it partially FT), which however requires discarding the circuit, resulting in sample loss. Similarly, the RGT allows only a single high-weight error to pass through, but the detection of errors requires preparing $|\overline{\theta}\rangle$ on the ancillary register in a repeat-until-success fashion without discarding the entire circuit. Therefore, the RGT implementation exhibits favorable scalability while retaining partial fault-tolerance.}

\subsection{Partially fault-tolerant encoding}
\label{sec:encoding}

We encode the circuit~\eqref{eq:qpe_circ} in terms of the Clifford+$R_{Z}$ gate set with the color code.
\begin{figure*}[htb]
    \centering
    \begin{minipage}{0.99\textwidth}
        \begin{quantikz}[style={column sep=2pt, row sep=7pt}]
            \qw
            & \qw
            & \ctrl{1}
            & \qw
            & \ctrl{1}
            & \gate{\mathrm{QEC}_X}
            & \qw
            &
            & \ctrl{1}
            & \qw
            & \ctrl{1}
            &
            & \qw
            & \qw
            & \gate{\mathrm{QEC}_{X,Z}}
            & \qw
            \\
            \qw
            & \gate[style={fill=green!15}]{\rgtQED_Z(h_{1}t)}
            & \gate{\overline{X}}
            & \gate[style={fill=green!15}]{\rgtQED_Z(-h_{1}t)}
            & \gate{\overline{X}}
            & \gate{\mathrm{QEC}_X}
            & \gate{\overline{H}}
            & \gate[style={fill=green!15}]{\rgtQED_Z(h_{2}t)}
            & \gate{\overline{X}}
            & \gate[style={fill=green!15}]{\rgtQED_Z(-h_{2}t)}
            & \gate{\overline{X}}
            & \gate{\overline{H}}
            & \qw
            & \qw
            & \gate{\mathrm{QEC}_{X,Z}}
            & \qw
        \end{quantikz}
    \end{minipage}
    \caption{Implementations of the encoded circuit component $\overline{{\rm ctrl-}U}$ introduced in Sec.~\ref{sec:encoding}. The top and bottom wires represent the ancilla and system registers, respectively. The Steane QEC gadgets, ${\rm QEC}_X$ for $X$ syndromes and ${\rm QEC}_Z$ for $Z$ syndromes, are given by \eqref{eq:circ_QEC_X} and \eqref{eq:circ_QEC_Z}, respectively.
    }
    \label{fig:circ_ctrlU}
\end{figure*}
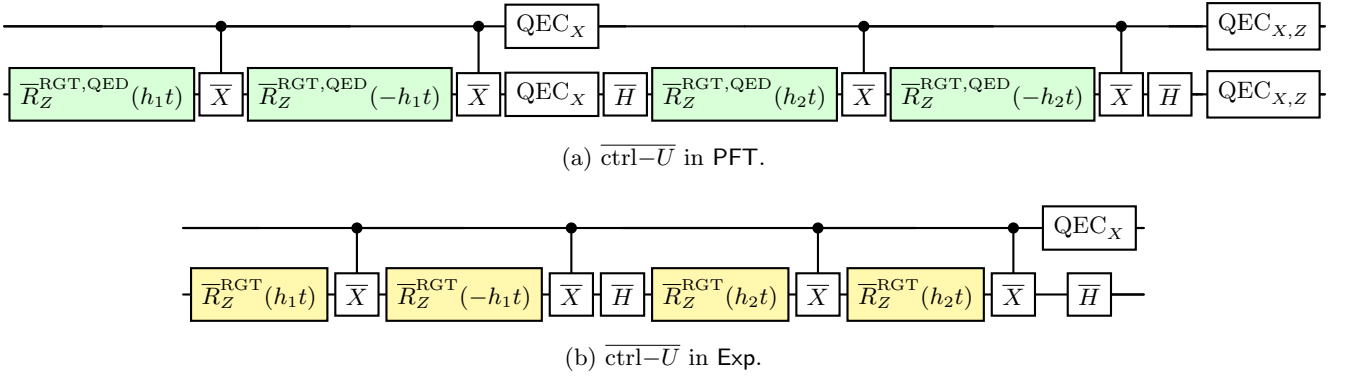
The logical circuit for preparing $\ket{\Phi}$ is compiled to
\begin{align}
\label{eq:circ_Phi_prep}
\begin{split}
    &\ket{\overline{\Phi}(\alpha_{0}, \alpha_{1})}
    =
    \\
    &\begin{quantikz}[style={column sep=4pt, row sep=8pt}]
        \lstick{$\ket{\overline{0}}$} & \gate{\overline{X}} &  \gate{\overline{H}} & \gate[style={fill=blue!10}]{\overline{R}_{Z}^{\mathrm{D}}(\alpha_{1})} & \gate{\overline{H}} & \gate{\overline{S}} & \gate[style={fill=blue!10}]{\overline{R}_{Z}^{\mathrm{D}}(\alpha_{2})} & \gate{\mathrm{QED}} & \qw \\
    \end{quantikz}
\end{split}
\end{align}
The circuit is discarded upon detecting errors by QED gadgets~\eqref{eq:circ_QED}~\cite{Self2022_natphys, Reichardt2024-jt}.
To encode $\overline{{\rm ctrl-}U}^k$, we decompose the unitary operator $U$~\eqref{eq:Trotter} in terms of the Clifford+$R_{Z}$ gate set. Each $\overline{R}_{Z}$ gate is implemented using the RGT gadget $\rgtQED_Z$.
As discussed in Sec.~\ref{subsec:qpe}, the evolution time is chosen such that $\e^{-\im h_{1}Zt} = T$.
Hence, $\rgtQED_Z$ for ctrl-$\e^{-\im h_{1}Zt}$~\cite{Trout2015-nr} is reduced to the standard $T$-gate teleportation.
To implement the ctrl-$\e^{-\im h_{2}Xt}$, we approximate $h_{2}t/\pi$ with a 5-bit binary fraction (Sec.~\ref{ssec:steane}), so that the recursion in RGT occurs up to 3 times. For each application of $\overline{{\rm ctrl-}U}$ depicted in Fig.~\ref{fig:circ_ctrlU}, we insert Steane QEC gadgets~\cite{Steane1997-tc,Postler2024-jc} with subscripts indicating the subset of stabilizers measured (see \eqref{eq:circ_QEC_X} and \eqref{eq:circ_QEC_Z}).
Finally, the phase-shift gate $\overline{R}_{Z}(\beta)$ for $\beta\in\{0,\pi/2\}$, which is Clifford, is implemented transversally.
{See Appendix~\ref{app:logical_components} for details of logical components in encoded circuits.}

\subsection{Calibration\label{ssec:calibration}}

To evaluate the accuracy of the QPE circuit execution, we prepare the eigenstate of $U$~\eqref{eq:Trotter} and set $\beta = -k\phi$ so that the measurement outcome is always zero without noise.
With hardware noise, the probability is reduced. We use a model of the measurement outcome probability with a free parameter~$q$ to account for the decoherence effects,
\begin{align}
\label{eq:P_calibration}
    P^{\rm (dec)}(q) = 1-\frac{q}{2}.
\end{align}
We model the decoherence effect $q$ by
\begin{align}
\label{eq:noise_q}
    q(k) = 1-\e^{-ak-b}%
\end{align}
where $a$ approximately represents the decay rate of the signal per ctrl-$U$ application.
We will analyze encoded circuits using this model, anticipating that it largely captures the influence of physical/logical errors on the circuit outcome.

We remark that since the phase-shift gate $R_Z(-k\phi)$ is generally not a Clifford gate, its logical gate can not be implemented transversally with the color code. Thus, we instead implement it non-fault-tolerantly with $\overline{R}^D_Z(-k\phi)$ in the calibration circuits.

\section{Experiment\label{sec:experiment}}

We experimentally demonstrate information theory QPE with quantum error correction on the Quantinuum H2-2~\cite{Moses2023-qw} after calibrating the QPE circuit~\eqref{eq:qpe_circ}.
Quantum circuits are implemented using \texttt{pytket} v1.38~\cite{Sivarajah2020-jg} and submitted with \texttt{qnexus} v0.10.0~\cite{quantinuum_nexus} to the hardware.
The physical circuit compilation is performed such that trivial redundancies (e.g., $H$ gates next to each other) are resolved and transpiled into the processor's native gate set.
We use extended \texttt{OpenQASM 2}~\cite{Cross2017-rz} for the lower-level representation.

As discussed in Sec.~\ref{sec:break-even}, the memory noise that accumulates during long idling and ion-transport times has an impact on computational accuracy.
To prevent the accumulation of coherent memory noise, which applies $\e^{-\im\varphi Z_i}$ on the idling $i$th qubit, we apply a variant of dynamical decoupling (DD)~\cite{Viola1999,Haghshenas2025}. The DD inserts a sequence of $X$ and $Y$ gates so that they do not alter logical operations but change the direction of the phase, e.g. $X\e^{-\im\varphi Z_i}X=\e^{\im\varphi Z_i}$, to prevent its accumulation.

\subsection{QPE circuit calibration}
\label{subsec:calibration}

To assess the utility of the Steane QEC gadgets in the encoded circuit (Sec.~\ref{sec:encoding}), we consider the settings with and without the Steane QEC gadgets.
For each setup and $k$, we run 500 shots of the QPE circuits to estimate the probabilities of obtaining $m=0$.
We fit the estimated probability with $P^{\rm (dec)}$~\eqref{eq:P_calibration} to read off the decoherence effect $q$, which gives a rough estimate of the circuit infidelity.

\begin{figure}[!htbp]
    \centering
    \includegraphics[width=0.99\linewidth]{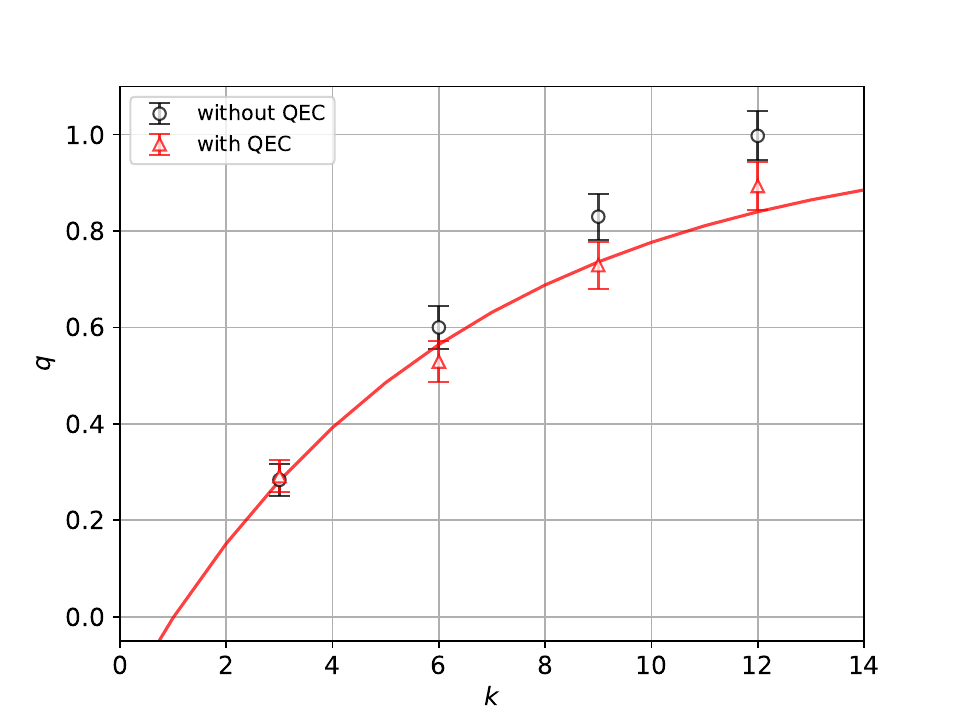}
    \caption{
        Decoherence parameter~$q$ obtained by fitting the probability of zero outcome from the QPE circuits with $P^{\rm (dec)}$~\eqref{eq:P_calibration}.
        The obtained $q$ is fitted with Eq.~\eqref{eq:noise_q} (solid curves).
        The probability is zero in the ideal case, leading to $q=0$.
    }
    \label{fig:calibration}
\end{figure}

As shown in Fig.~\ref{fig:calibration}, the encoded circuit with QEC gadgets shows a smaller decoherence effect $q(k=12)$ than the one without the Steane QECs. This is attributed to the real-time Steane QECs preventing errors from building up, particularly at large $k$.
Using Eq.~\eqref{eq:noise_q}, the error parameters are estimated to be $a = 0.17(2)$ and $b = -0.17(9)$.

\subsection{Ground state energy estimation\label{sec:energy}}

We proceed to the ground state energy estimation following Sec.~\ref{subsec:qpe}.
We randomly draw $\Ns/10$ samples of $(k, \beta)$ with $k_{\mathrm{max}} = 12$, and take $10$ shots of circuit executions for each pair. The QED gadgets discard $4.7\%$ of the total shots, which are excluded from the number of samples $\Ns$.
To account for the hardware noise in post-processing, we use the noise-aware distribution,
\begin{align}
\label{eq:probs_single_noise}
    Q(m_l|\phiest; k_l,\beta_l)
    =
    \frac{1 + \e^{-ak-b}\cos(k_l\phiest + \beta_l - m_l\pi)}{2}
\end{align}
instead of Eq.~\eqref{eq:probs_single}, where $a$ and $b$ are obtained from the calibration in Sec.~\ref{subsec:calibration}.

\begin{figure}[!htbp]
    \centering
    \includegraphics[width=\linewidth]{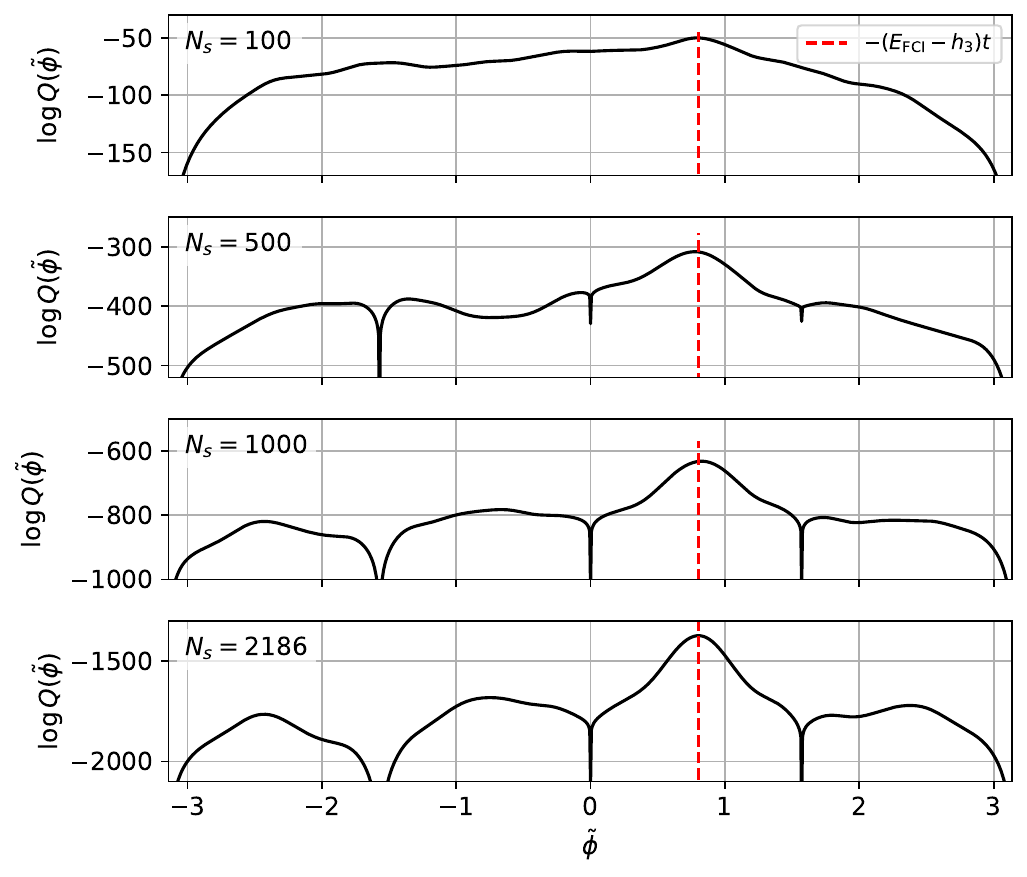}
    \caption{
        The distributions [Eqs~\eqref{eq:probs} and \eqref{eq:probs_single_noise}] obtained by experimentally collecting the measurement outcomes of the circuit Eq.~\eqref{eq:qpe_circ} for different numbers of samples~$\Ns$.
        The target phase, $-E_{\rm FCI}\,t$, is indicated by the red dashed line.
        \label{fig:iqpe}
    }
\end{figure}

The experimentally obtained distributions [Eqs~\eqref{eq:probs} and \eqref{eq:probs_single_noise}] over the phase $\phiest$ are plotted for various $N_{\mathrm{s}}$ in Fig.~\ref{fig:iqpe}.
As $\Ns$ is increased, the dominant peak of the distribution becomes sharper around the target phase $-E_{\rm FCI}\,t$, and thus, the estimate of ground state energy becomes more accurate.
From the distribution, we estimate the energy and statistical uncertainty as follows. We draw a sample of size $\Ns$ with replacement from the data $\{m_l\}_{l=1}^{\Ns}$ and compute the phase $\phi^{(i)}$ with~\eqref{eq:argmax}. After repeating it $R$ times to obtain the collection $\{\phi^{(i)}\}_{i=1}^{R}$, we estimate the phase by the circular mean $\mathrm{Arg}(\sum_{i}\e^{\im\phi^{(i)}}/R)$. The uncertainty is evaluated from the Holevo variance $|\sum_{i=1}^{R}\e^{\im\phi^{(i)}}/R|^{-2}-1$.
With $R=1000$, the energy is estimated as $E = -1.136(13)$~hartree after collecting $\Ns=2186$ samples, which agrees with the FCI energy $E_{\mathrm{FCI}} = -1.13731$ hartree within the statistical uncertainty of $\sigma$.

While the error of the estimated energy is still larger than the chemical precision ($\sim0.0016$ hartree), we emphasize that our end-to-end pipeline will be readily improved for more accurate energy estimation as hardware capabilities continue to advance, such as increases in qubit count that enable higher-distance codes.
\section{Numerical simulations with tunable noise parameters\label{sec:numerical}}

We investigate the impact of hardware noise on a partially fault-tolerant QPE circuit via numerical simulations with tunable noise parameters.
The simulations presented in this section are performed with the H2 emulator.

\subsection{Noise model}
\label{ssec:noise_model}

We focus on the three dominant sources of hardware noise: those coming from two-qubit gates and measurements, and memory noise~\cite{Moses2023-qw}.
All the other noise sources, such as those coming from single-qubit gates, are ignored in the simulation to examine the influence of the three main sources exclusively. The noise models are summarized in Tab.~\ref{tab:noise_model}.
\begin{table}[!htbp]
    \centering
    \def\arraystretch{1.5}
    \begin{tabular}{l|c||c}
        \hline
        \multicolumn{2}{c||}{Noise model} & Parameter
        \\
        \hline\hline
        \multicolumn{2}{c||}{two-qubit-gate fault} & $p_2$
        \\
        \hline
        \multirow{2}{*}{Readout fault}& (``1''$\to$``0'') & $p_{\rm r,1\to0}$
        \\
        \cline{2-3}
        & (``0''$\to$``1'') & $p_{\rm r,0\to1}$
        \\
        \hline
        \multirow{2}{*}{Memory noise} & coherent & $f$
        \\
        \cline{2-3}
        & incoherent & $g$
        \\
        \hline
    \end{tabular}
    \caption{\label{tab:noise_model}
    Noise models used in our numerical simulations. The parameters for the two-qubit gate and readout faults stand for the probabilities of the respective faults occurring. The parameters $f$ and $g$ characterize coherent and incoherent memory noises~\eqref{eq:memory_noise}.
    }
\end{table}

The two-qubit gate fault is modeled by an anisotropic two-qubit Pauli channel.
The readout fault flips a result from ``1'' to ``0'' with the probability $p_{\rm r,1\to0}$, and from ``0'' to ``1'' with the probability $p_{\rm r,0\to1}$.
The memory noise on idling or moving qubits is modeled with the following dephasing channel:
\begin{align}
\label{eq:memory_noise}
    \rho
    \mapsto
    \e^{-\im ft_{\rm m}Z/2}((1-gt_{\rm m})\rho + gt_{\rm m} Z\rho Z) \e^{\im ft_{\rm m}Z/2},
\end{align}
where $t_m$ is the idling/transport time of the qubit on which the channel acts. The parameters $f$ and $g$ characterize the coherent and incoherent parts of memory noise.
See \cite{H2emulator} for details of the noise model.

\subsection{Exploring dominant noise sources\label{sec:break-even}}

To identify the dominant noise source under the partially FT setup, we classically simulate the encoded QPE circuits and study separately the effects of two-qubit-gate and readout fault, and coherent and incoherent memory noise (Table.~\ref{tab:data}).
The H2-2 quantum computer is emulated with the parameters $p_2=1.29\times10^{-3}$, $f=4.3\times10^{-2}$, and $g=2.8\times10^{-3}$~\cite{H2emulator}.

\begin{table}[!htbp]
\centering
\def\arraystretch{1.5}
\begin{tabular}{c||c|c}
\hline
  Noise type
  & \quad $q(k=6)$ \enspace & \quad $q(k=12)$ \enspace
\\\hline\hline
  Gate \& readout error & $0.11(2)$ & $0.19(5)$  \\\hline
  Coherent memory noise & $0.000(0)$ & $0.012(7)$  \\\hline
  Incoherent memory noise & $0.27(3)$ & $0.63(4)$  \\\hline
\end{tabular}
\caption{\label{tab:data}
    Data of noisy encoded circuit simulation with different noise sources turned on. The first row is obtained with the noise parameters $p_2=1.29\times 10^{-3}$, $p_{\rm r, 0\to1}=0.9\times 10^{-3}$, and $p_{\rm r,1\to0}=1.8\times 10^{-3}$. The second and third rows are calculated with the coherent memory noise $f=4.3\times 10^{-2}$[rad$\cdot$ s$^{-1}$] and incoherent memory noise $g=2.8\times 10^{-3}$[s$^{-1}$], respectively. For each setup, all the unspecified noises are turned off.
}
\end{table}

\begin{table}[!htbp]
{
\centering
\def\arraystretch{1.5}
\begin{tabular}{c||c}
\hline
  Operation type
  & Execution time (fraction)
\\\hline\hline
  $\overline{CX} \times 4$ & 2\%  \\\hline
  $\rgtQED_Z(h_1 t)$ \& $\rgtQED_Z(-h_1 t)$ & 20\%  \\\hline
  $\rgtQED_Z(h_2 t)$ \& $\rgtQED_Z(-h_2 t)$ & 56\%  \\\hline
  QEC$_X \times 4$ \& QEC$_Z \times 2$ &  22\% \\\hline
\end{tabular}
\caption{\label{tab:data_time}
    Breakdown of typical operation time per $\overline{{\rm ctrl-}U}$ in Fig.~\ref{fig:circ_ctrlU}. Since $h_1t=\pi/8$, $R_Z(\pm h_1t)$ gates are reduced to $T/T^\dag$ gates, resulting in the short execution time relative to that of $R_Z(\pm h_2t)$. The execution time for $\overline{H}$ is not reported as it is negligible.
}
}
\end{table}

Table~\ref{tab:data} implies that the incoherent memory noise is the dominant noise source under the present setup.
The encoded circuit incurs long idling time mainly due to intricately nested conditional operations in RGTs and QEC gadgets, leading to a large impact of memory noise. {See Tab.~\ref{tab:data_time} for the allocation of quantum operation times within a single $\overline{{\rm ctrl-}U}$ gate.}
Nevertheless, the second row of Tab.~\ref{tab:data} shows that the coherent memory noise is well controlled even for the current value, $f=4.3\times10^{-2}\text{rad}\cdot\text{s}^{-1}$. This is attributed to the use of DD, which alleviates the accumulation of coherent memory noise.
As a matter of fact, the decoherence effect $q(k=6,12)$ is nearly 1 without DD for $f=10^{-1.5}\text{rad}\cdot\text{s}^{-1}$ and $p_2=g=0$.
While periodic applications of the QEC gadgets associated with $X$ stabilizers provide protection against the noise, lack of other error suppression mechanisms makes it harder to control the incoherent part of memory noise.
For reference, the values $q(k=6,12)$ with and without DD coincide within statistical errors for $g=10^{-2.5}\text{s}^{-1}$ and $p_2=f=0$, implying the DD does not suppress the incoherent memory noise effect as expected.

\section{Conclusions and outlook\label{sec:conclusions}}

We have experimentally demonstrated an end-to-end workflow to compute molecular energies.
We have performed the information theory QPE algorithm encoded with the color code to estimate the ground state energy of the hydrogen molecule in minimal basis using a trapped-ion quantum computer.
In the calibration of QPE circuits, we showed that the introduction of QEC gadgets can suppress noise, rather than proliferate it, despite additional circuit overhead. This encouraging observation implies that intermediate QECs are necessary to enable deeper and higher-fidelity circuit executions.
In the ground-state energy estimation, the error in the estimate is $0.001(13)$ hartree with respect to the exact value within the given basis set.
Though the estimated energy error is still larger than the chemical precision ($\sim0.0016$ hartree), this is a critical achievement towards fully fault-tolerant quantum computational chemistry simulations.

Numerical simulation on an emulator with tunable noise parameters suggests that incoherent memory noise is the main source of error for encoded circuits.
Developments oriented to increase the protection against memory noise are consequently deemed to be the most effective in improving the experimental results.
We note that all experiments and simulations were performed using a small $d=3$ code, which is only able to detect and correct a \textit{single} error.
Indeed, increasing the code distance allows for more robust computation and the execution of longer circuits, like QPE, as it suppresses all sources of noise, including memory noise.
Increasing the distance could either be done with an increase in the number of qubits, or the use of a higher-rate qLDPC code~\cite{Roffe2023}.
Furthermore, as the hardware noise is biased toward dephasing, using QEC codes that provide stronger protection against Pauli-$Z$ errors~\cite{Dasu2025, Dua2024, Roffe2023, Monz2009}, like rectangular surface codes~\cite{Tuckett2019}, can provide additional protection.

There are also circuit techniques beyond DD that could be implemented, such as Pauli twirling~\cite{cai2020mitigating}, Clifford deformation~\cite{dua2024clifford}, or optimizing stabilizer parities~\cite{debroy2021optimizing}, which could further tailor the code to the noise of the environments.
On the logical circuit level, optimized QPE algorithms~\cite{dutkiewicz_arxiv2025, vonBurg2021, Lee2021, Paul2025-gz, caesura_arxiv2025}, or more general advanced architecture designs~\cite{Gidney2021, Litinski2019} and compilation strategies~\cite{Hirano2025, Watkins2024} minimize the space-time volume where memory noise accumulates.

We also expect better compilation techniques—those that compile on the logical level rather than on the physical level and are optimized for QEC protocols—will help reduce the noise a circuit experiences.
Finally, compilation languages beyond extended \texttt{OpenQASM2} or the more sophisticated quantum intermediate representation (\texttt{QIR}) can reduce the size of the compiled circuit and prevent the accumulation of memory noise.
Exploration of these techniques is left for future work.

The present results demonstrate that current quantum hardware, in conjunction with quantum error correction protocols, can support scalable quantum computational chemistry calculations and yield meaningful energy estimates.
We believe this work marks an important step toward fault-tolerant quantum computational chemistry and will motivate continued theoretical and experimental efforts to further advance the field.

\section*{Resources}

The source codes used in the present experiments and numerical simulations are available from \href{https://github.com/CQCL/qec_qpe_chem}{\texttt{https://github.com/CQCL/qec\_qpe\_chem}}.

\section*{Acknowledgments}

We are grateful to the entire Quantinuum team for their many contributions to the work. We thank Charles~H.~Baldwin,
Eli~Chertkov,
Regina~Finsterhoelzl,
Michael Foss-Feig,
David Hayes,
Andrew Potter,
Grahame Vittorini, and
James Walker
for useful discussions and feedback on the manuscript.

\bibliography{main}

\onecolumngrid
\clearpage
\appendix
\renewcommand\thefigure{\thesection \arabic{figure}}
\renewcommand\thetable{\thesection \arabic{table}}
\onecolumngrid
\section{Information theory QPE}
\label{app:ipqe}

We discuss the information theory QPE~\cite{Svore2014-ka} to supplement Sec.~\ref{subsec:qpe}.
We show that the distribution~\eqref{eq:probs},
\begin{align}
\label{eq:app_probs}
    Q(\phiest\mid \{m_{l}\}; \{k_{l},\beta_{l}\})
    =
    \prod_{l=1}^{N_{\mathrm{s}}}
    Q(m_{l}|\phiest; k_{l},\beta_{l})
    \quad \text{with} \quad
    Q(m_l|\phiest; k_l,\beta_l)
    =
    \frac{1 + \cos(k_l\phiest + \beta_l - m_l\pi)}{2},
\end{align}
has peaks around the eigenphases $\{\phi_j\}_j$.
To this end, we calculate
\begin{align}
\label{eq:average_probs}
    \bE_{\{m_l,\beta_l,k_l\}}\big[
        Q(\phiest|\{m_{l}\}; \{k_{l},\beta_{l}\})
    \big],
\end{align}
where the average over $m_l$ is taken according to the probablity~\eqref{eq:prob_circ}.
The averages over $\beta_l$ and $k_l$ are taken uniformly over $[0,2\pi)$ and $\{1,2,\dots,k_{\rm max}\}$, respectively.
Since $\Ns$ experiments are independent, the probablity~\eqref{eq:average_probs} can be factorized as
\begin{align}
    \bE_{\{m_l,\beta_l,k_l\}}\big[
        Q(\{m_{l}\}|\phiest; \{k_{l},\beta_{l}\})
    \big]
    =
    \big(\bE_{m,\beta,k}\big[
        Q(m|\phiest; k,\beta)
    \big]\big)^{\Ns}.
\end{align}
The average for each experiment is calculated as,
\begin{align}
\begin{split}
    \bE_{m,\beta,k}\big[
        Q(m|\phiest; k,\beta)
    \big]
    &=
    \sum_{m=0,1}\bE_{\beta,k}\big[
        Q(m|\phiest; k,\beta)P(m|\{a_j, \phi_j\}; k,\beta)
    \big]
    \\
    &=
    \sum_{j}|a_j|^2\bE_{\beta,k}\bigg[
        \frac{1 + \cos(k\phi_j + \beta)\cos(k\phiest + \beta)}{2}
    \bigg]
    \\
    &=
    \frac{1}{2} + \frac{1}{4}\sum_{j}|a_j|^2\bE_{k}
    [\cos(k(\phi_j-\phiest))].
\end{split}
\end{align}
Letting $\delta_j:=\phi_j-\phiest$, we calculate $\bE_{k}[\cos(k(\phi_j-\phiest))]$,
\begin{align}
    \bE_{k}[\cos(k\delta_j)]
    =
    \frac{1}{k_{\rm max}}\sum_{k=1}^{k_{\rm max}}\cos(k\delta_j)
    =
    \frac{\sin(k_{\rm max}\delta_j/2)\cos((k_{\rm max}+1)\delta_j/2)}{k_{\rm max}\sin(\delta_j/2)}.
\end{align}
Thus, we arrive at
\begin{align}
\begin{split}
    \bE_{m,\beta,k}\big[
        Q(m|\phiest; k,\beta)
    \big]
    &=
    \frac{1}{2} + \frac{1}{4}\sum_{j}|a_j|^2\frac{\mathrm{sinc}(k_{\rm max}\delta_j/2)\cos((k_{\rm max}+1)\delta_j/2)}{k_{\rm max}\mathrm{sinc}(\delta_j/2)},
\end{split}
\end{align}
where $\mathrm{sinc}(\alpha x):=\sin(\alpha x)/\alpha x$ has the peak at $x=0$ of height 1 and width $\sim\alpha^{-1}$.
For $\phiest$ away from all $\phi_j$ ($\delta_j\gg 1/k_{\rm max}$ for all $j$),
\begin{align}
\begin{split}
    \bE_{m,\beta,k}\big[
        Q(m|\phiest; k,\beta)
    \big]
    &\approx
    \frac{1}{2}.
\end{split}
\end{align}
For $\phiest$ in the vicinity of $\phi_j$ ($k_{\rm max}\delta_j\ll 1$ for some $j$), we have
\begin{align}
\begin{split}
    \bE_{m,\beta,k}\big[
        Q(m|\phiest; k,\beta)
    \big]
    &\approx
    \frac{2+|a_j|^2}{4}\e^{-(k_{\rm max}\delta_j)^2/(6+12/A_j)}.
\end{split}
\end{align}

\begin{figure}[!htbp]
    \centering
    \begin{minipage}{0.49\textwidth}
        \includegraphics[width=\linewidth]{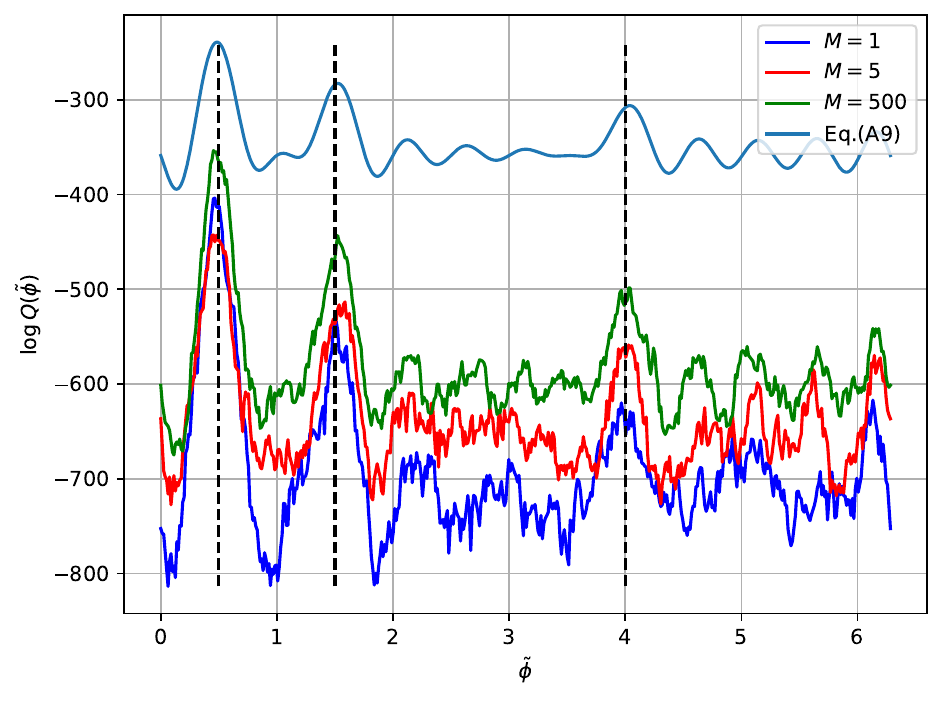}
    \end{minipage}
    \begin{minipage}{0.49\textwidth}
        \includegraphics[width=\linewidth]{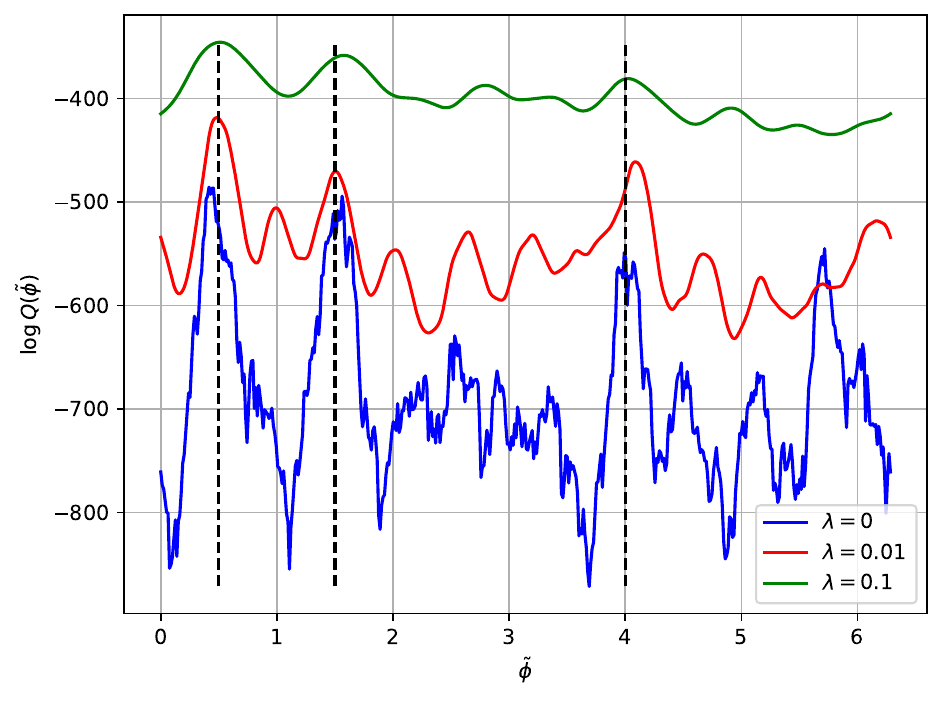}
    \end{minipage}
    \caption{\label{fig:iqpe_app}
        (Left panel) Comparison between the analytic expression~\eqref{eq:average_prob_expression} and the distributions obtained by sampling measurement outcomes according to Eq.~\eqref{eq:prob_circ}. For blue, red, and green data, we collect $M$ sets of measurement outcomes $\{m_l\}_{l=1}^{\Ns}$, calculate the distribution~\eqref{eq:app_probs} for each set, and then average $M$ distributions.
        (Right panel) Impact of hardware noise on the noise-aware distribution, \eqref{eq:app_probs} and \eqref{eq:probs_single_noise}, obtained by sampling measurement outcomes according to Eq.~\eqref{eq:prob_noisy_circ}. Here, $M$ is fixed to $1$.
        For both panels, we use the input state~\eqref{eq:trial_state} with $(|a_0|^2,|a_1|^2,|a_2|^2)=(0.5,0.3,0.2)$ and $(\phi_0,\phi_1,\phi_2)=(0.5,1.5,4)$, indicated by the vertical dashed lines, and choose $\Ns=500$ and $k_{\rm max}=12$.
    }
\end{figure}

Therefore, upon averaging over $\{m_l,\beta_l,k_l\}$, the distribution~\eqref{eq:average_probs} takes the form,
\begin{align}
\label{eq:average_prob_expression}
    \bE_{\{m_l,\beta_l,k_l\}}\big[
        Q(\phiest|\{m_{l}\}; \{k_{l},\beta_{l}\})
    \big]
    =
    \bigg(\frac{1}{2}+\frac{1}{4}\sum_j A_j
        \frac{\mathrm{sinc}(k_{\rm max}\frac{\phiest-\phi_j}{2})\cos((k_{\rm max}+1)\frac{\phiest-\phi_j}{2})}{\mathrm{sinc}(\frac{\phiest-\phi_j}{2})}
    \bigg)^{\Ns},
\end{align}
which is peaked around $\phi_j$ with the height $((|a_j|^2+2)/4)^{\Ns}$ and width $\sim 1/(k_{\rm max}\sqrt{\Ns |a_j|^2})$.

The left panel of Fig.~\ref{fig:iqpe_app} compares the analytic expression~\eqref{eq:average_prob_expression} and the distributions obtained by sampling measurement outcomes according to Eq.~\eqref{eq:prob_circ}.
We collect $M$ sets of measurement outcomes $\{m_l\}_{l=1}^{\Ns}$, calculate the distribution~\eqref{eq:app_probs} for each set, and then average $M$ distributions. As $M$ is increased, the distribution approaches the mean distribution~\eqref{eq:average_prob_expression}.
For all the data, peaks are found near the eigenphases, with the dominant one around $\phi_0$.
The hardware experiment presented in Sec.~\ref{sec:energy} is done with $M=1$.

The right panel of Fig.~\ref{fig:iqpe_app} shows the influence of hardware noise on the distribution by sampling the measurement outcomes from
\begin{align}
\label{eq:prob_noisy_circ}
    P(m|\{A_j, \phi_j\}; k,\beta)
    =
    \sum_{j=0}
    |a_j|^2
    \frac{1 + \e^{-\lambda k}\cos(k\phi_{j} + \beta - m\pi)}{2}
\end{align}
instead of Eq.~\eqref{eq:prob_circ}. Here, we employ the noise-aware distribution~\eqref{eq:probs_single_noise} to calculate~\eqref{eq:app_probs}.
The noise smears the distribution, and thus, it requires more shots to accurately locate the peak positions using Eq.~\eqref{eq:argmax}.

\section{Logical components}
\label{app:logical_components}

We collect all the logical components in the encoded quantum circuits used in the main text.

\subsection{Clifford operations}

The logical Clifford gates $\{\overline{S}$, $\overline{H}, \overline{CX}\}$ admitted by the color code can be implemented transversally.
\begin{align}
\label{eq:clifford}
    \overline{H} = H^{\otimes 7},
    \qquad
    \overline{S} = S^{\dagger\otimes 7},
    \qquad
    \overline{CX} =
    \begin{quantikz}[style={column sep=4pt, row sep=4pt}]
        \lstick{} & \ctrl{1} & \qw \\
        \lstick{} & \gate{\overline{X}} & \qw \\
    \end{quantikz}
    ~=~
    \begin{quantikz}[style={column sep=4pt, row sep=4pt}]
        \lstick[7]{block$_1$} & \ctrl{7} & \qw      & \qw      & \qw      & \qw      & \qw      & \qw      & \qw \\
        & \qw      & \ctrl{7} & \qw      & \qw      & \qw      & \qw      & \qw      & \qw \\
        & \qw      & \qw      & \ctrl{7} & \qw      & \qw      & \qw      & \qw      & \qw \\
        & \qw      & \qw      & \qw      & \ctrl{7} & \qw      & \qw      & \qw      & \qw \\
        & \qw      & \qw      & \qw      & \qw      & \ctrl{7} & \qw      & \qw      & \qw \\
        & \qw      & \qw      & \qw      & \qw      & \qw      & \ctrl{7} & \qw      & \qw \\
        & \qw      & \qw      & \qw      & \qw      & \qw      & \qw      & \ctrl{7} & \qw \\
        \lstick[7]{block$_2$} & \targ{}  & \qw      & \qw      & \qw      & \qw      & \qw      & \qw      & \qw \\
        & \qw      & \targ{}  & \qw      & \qw      & \qw      & \qw      & \qw      & \qw \\
        & \qw      & \qw      & \targ{}  & \qw      & \qw      & \qw      & \qw      & \qw \\
        & \qw      & \qw      & \qw      & \targ{}  & \qw      & \qw      & \qw      & \qw \\
        & \qw      & \qw      & \qw      & \qw      & \targ{}  & \qw      & \qw      & \qw \\
        & \qw      & \qw      & \qw      & \qw      & \qw      & \targ{}  & \qw      & \qw \\
        & \qw      & \qw      & \qw      & \qw      & \qw      & \qw      & \targ{}  & \qw \\
    \end{quantikz}
\end{align}
where $H^{\otimes 7}$ and $S^{\dag\otimes 7}$ act on all the qubits transversally in a single code block, and block$_i$ in the $\overline{CX}$ circuit represents a code block.

\subsection{Preparation of \texorpdfstring{$\ket{\overline{0}}$}{the logical zero state}}

We summarize (non-)FT circuits to encode the state $\ket{\overline{0}}$.

\begin{itemize}

\item Non-FT preparation of $\ket{\overline{0}}$

The logical state $\ket{\overline{0}}$ is non-fault-tolerantly prepared by the following circuit:
\begin{align}
\label{eq:circ_prep_nonFT}
    \ket{\overline{0}^{\rm nonFT}}
    ~=~
    \begin{quantikz}[style={column sep=4pt, row sep=5pt}]
        \lstick{$q_0$: $\ket{0}$} &  \gate[1]{\text{$H$}} & \ctrl{1}  & \ctrl{3} & \qw       & \qw       & \qw        & \qw \\
        \lstick{$q_1$: $\ket{0}$} &  \qw                  & \targ{}   & \qw      & \qw       & \targ{}   & \qw        & \qw \\
        \lstick{$q_2$: $\ket{0}$} &  \qw                  & \qw       & \qw      & \targ{}   & \qw       & \targ{}    & \qw \\
        \lstick{$q_3$: $\ket{0}$} &  \qw                  & \targ{}   & \targ{}  & \qw       & \qw       & \ctrl{-1}  & \qw \\
        \lstick{$q_4$: $\ket{0}$} &  \gate[1]{\text{$H$}} & \qw       & \ctrl{1} & \ctrl{-2} & \ctrl{-3} & \qw        & \qw \\
        \lstick{$q_5$: $\ket{0}$} &  \qw                  & \qw       & \targ{}  & \targ{}   & \qw       & \qw        & \qw \\
        \lstick{$q_6$: $\ket{0}$} &  \gate[1]{\text{$H$}} & \ctrl{-3} & \qw      & \ctrl{-1} & \qw       & \qw        & \qw \\
    \end{quantikz}
\end{align}
where $q_i$ represents the $i$$^\mathrm{th}$ physical qubit comprising one logical qubit of the color code. This non-FT circuit is used in the experiment presented in App.~\ref{app:non_ft}.
Note that we use $\ket{\overline{0}^{\rm nonFT}}$ to represent the non-fault-tolerant circuit to prepare the state $\ket{\overline{0}}$. We adopt the same convention for the superscript ``FT'' (fault-tolerant).

\item FT preparation of $\ket{\overline{0}}$

A FT preparation of the state $\ket{\overline{0}}$ is accomplished by the circuit~\cite{Goto2016-rl}:
\begin{align}
\label{eq:circ_prep_FT}
    \ket{\overline{0}^{\mathrm{FT}}}
    ~=~
    \begin{quantikz}[style={column sep=4pt, row sep=5pt}]
        \lstick{$q_0$: $\ket{0}$} &  \gate[1]{\text{$H$}} & \ctrl{1}  & \ctrl{3} & \qw       & \qw       & \qw        & \qw      & \qw      & \qw      & \qw & \qw \\
        \lstick{$q_1$: $\ket{0}$} &  \qw                  & \targ{}   & \qw      & \qw       & \targ{}   & \qw        & \ctrl{6} & \qw      & \qw      & \qw & \qw \\
        \lstick{$q_2$: $\ket{0}$} &  \qw                  & \qw       & \qw      & \targ{}   & \qw       & \targ{}    & \qw      & \qw      & \qw      & \qw & \qw \\
        \lstick{$q_3$: $\ket{0}$} &  \qw                  & \targ{}   & \targ{}  & \qw       & \qw       & \ctrl{-1}  & \qw      & \ctrl{4} & \qw      & \qw & \qw \\
        \lstick{$q_4$: $\ket{0}$} &  \gate[1]{\text{$H$}} & \qw       & \ctrl{1} & \ctrl{-2} & \ctrl{-3} & \qw        & \qw      & \qw      & \qw      & \qw & \qw \\
        \lstick{$q_5$: $\ket{0}$} &  \qw                  & \qw       & \targ{}  & \targ{}   & \qw       & \qw        & \qw      & \qw      & \ctrl{2} & \qw & \qw \\
        \lstick{$q_6$: $\ket{0}$} &  \gate[1]{\text{$H$}} & \ctrl{-3} & \qw      & \ctrl{-1} & \qw       & \qw        & \qw      & \qw      & \qw      & \qw & \qw \\
        \lstick{$a_{0}$: $\ket{0}$} &  \qw                  & \qw       & \qw      & \qw       & \qw       & \qw        & \targ{}  & \targ{}  & \targ{}  & \meter{} & \rstick{$c\in\{0,1\}$}\cw \\
    \end{quantikz}
\end{align}
where $a_0$ represents the ancillary qubit.
The circuit first encodes the state $\ket{\overline{0}}$ non-fault-tolerantly with \eqref{eq:circ_prep_nonFT} and then measures the bottom qubit to check if the state is correctly prepared up to a weight-1 $X$ error. A successful encoding is indicated by $c=0$. The circuit is discarded when $c=1$ is measured, and the protocol repeats the circuit executions until $c=0$ is measured.

\end{itemize}

\subsection{Measurement}
\label{app:circ_meas}

The quantum circuit to destructively measure a logical Pauli $\overline{P}\in\{\overline{X},\overline{Z}\}$ is given by
\begin{align}
\label{eq:circ_meas}
    \begin{quantikz}[style={column sep=6pt, row sep=5pt}]
        \lstick{} & \meterD{\overline{P}}
    \end{quantikz}
    =
    \begin{quantikz}[style={column sep=4pt, row sep=5pt}]
        & \qw  & \meterD{P} & \setwiretype{n} \cw & \rstick {$c_{0}$} \\
        & \qw  & \meterD{P} & \setwiretype{n} \cw & \rstick {$c_{1}$} \\
        & \qw  & \meterD{P} & \setwiretype{n} \cw & \rstick {$c_{2}$} \\
        & \qw  & \meterD{P} & \setwiretype{n} \cw & \rstick {$c_{3}$} \\
        & \qw  & \meterD{P} & \setwiretype{n} \cw & \rstick {$c_{4}$} \\
        & \qw  & \meterD{P} & \setwiretype{n} \cw & \rstick {$c_{5}$} \\
        & \qw  & \meterD{P} & \setwiretype{n} \cw & \rstick {$c_{6}$} \\
    \end{quantikz}
\end{align}
From the seven bits of measurement outcome $\{c_i\}$, we extract the syndromes $s_1, s_2, s_3$, associated to the stabilizers $S_{X0}, S_{X1}, S_{X2}$ in Fig.~\ref{fig:steane}, to identify and correct a weight-one error.
Then, we evaluate the logical readout $\overline{c}$,
\begin{align}
    \label{eq:decode}
    \overline{c} = c_1 \oplus c_3 \oplus c_5,
\end{align}
with the corrected bits.

\subsection{QEC and QED gadgets}

\begin{itemize}

\item Steane QEC gadgets

The Steane QEC gadget using all the $X$ stabilizers is given by,
\begin{align}
\label{eq:circ_QEC_X}
    \begin{quantikz}[style={column sep=4pt, row sep=5pt}]
        \lstick{} & \qw & \gate{\mathrm{QEC}_X} & \qw \\
    \end{quantikz}
    ~=~
    \begin{quantikz}[style={column sep=4pt, row sep=5pt}]
        \lstick{} \qw                                      & \gate{\overline{X}}   & \qw & \gate{Z_\mathrm{EC}}         & \qw \\
        \lstick{$\ket{\overline{0}^{\mathrm{FT}}}$}    \qw & \ctrl{-1} & \qw & \meterD{\overline{X}} \vcw{-1} \\
    \end{quantikz}
\end{align}
The gate \begin{quantikz}[{column sep=2pt}]&\gate{Z_{\rm EC}}&\end{quantikz} represents the $Z$ gates that correct the detected $Z$ errors.
Similarly, the Steane QEC gadget using the $Z$ stabilizers is given by
\begin{align}
\label{eq:circ_QEC_Z}
    \begin{quantikz}[style={column sep=4pt, row sep=5pt}]
        \lstick{} & \qw & \gate{\mathrm{QEC}_Z} & \qw \\
    \end{quantikz}
    ~=~
    \begin{quantikz}[style={column sep=4pt, row sep=5pt}]
        \lstick{} & \qw                                      & \ctrl{1} & \qw & \gate{X_\mathrm{EC}}         & \qw \\
        \lstick{$\ket{\overline{+}^{\mathrm{FT}}}$}    & \qw & \gate{\overline{X}}  & \qw & \meterD{\overline{Z}} \vcw{-1} \\
    \end{quantikz}
\end{align}

{
The FT preparation of $\ket{\overline{+}^{\mathrm{FT}}}$ is repeated until success as given in Algorithm~\eqref{alg:QEC}.
In the experiment, we set the maximum number of repetitions $R_{\rm max}=2$, i.e., the circuit execution is aborted if the initialization fails $R_{\rm max}$ times.
}

\begin{algorithm}
{
\caption{Steane QEC$_X$ gadget}
\SetKwInOut{Input}{input}
\SetKwInOut{Output}{output}
\SetKwInOut{Registers}{registers}
\SetKw{Run}{Run}
\label{alg:QEC}
    \Input{
        the maximum number of recursions $r_{\rm rus}\in\mathbb{N}$.
    }
        \Registers{
        Input state on system register $\mathtt{QREG_{\rm sys}}$,
        Ancilla register $\mathtt{QREG_{\rm anc}}$.
    }
    $r\gets0$\;
    \While{$r<r_\mathrm{rus}$}{
        \Run{the circuit~\eqref{eq:circ_prep_FT} on $\mathtt{QREG_{\rm anc}}$}\;
        \If{no error is detected in \eqref{eq:circ_prep_FT}}{
            \Run{the circuit~\eqref{eq:circ_QEC_X} on $(\mathtt{QREG_{\rm sys}},\mathtt{QREG_{\rm anc}})$} \tcp*{returns an error-corrected state on $\mathtt{QREG_{\rm sys}}$}
            \Output{$\mathtt{QREG_{\rm sys}}$}
        }
        $r\gets r+1$\;
    }
    Abort the circuit execution \tcp*{Discard the shot}
    \Output{failure}
}
\end{algorithm}

\item FT QED gadgets

The FT QED gadget using the $S_{X0}, S_{Z0}$ stabilizers is given by
\begin{align}
    \begin{quantikz}[style={row sep=5pt, column sep=4pt}]
        \lstick{}          & \gate{{\rm QED}_0} & \qw \\
    \end{quantikz}
    ~=~
    \begin{quantikz}[style={row sep=8pt, column sep=4pt}]
        \lstick{$q_0$ \quad\ \ }
        &
        & \targ{}   & \ctrl{4}
        &           &
        & \qw       & \qw
        & \qw       & \qw
        & \qw       & \qw
        &
        \\
        \lstick{$q_1$ \quad\ \ }
        &
        &           &
        & \ctrl{3}  & \targ{}
        &           &
        &           & \qw
        &           & \qw
        &
        \\
        \lstick{$q_3$ \quad\ \ }
        &
        &           &
        &           &
        & \targ{}   & \ctrl{2}
        & \qw       & \qw
        & \qw       & \qw
        &
        \\
        \lstick{$q_4$ \quad\ \ }
        &
        &           &
        &           &
        &           &
        & \ctrl{1}  & \targ{}
        &           &
        &
        \\
        \lstick{$a_{0}: \ket{0}$}
        &
        &           & \targ{}
        & \targ{}   &
        &           & \targ{}
        & \targ{}   &
        &           & \meter{}
        & \cw
        \\
        \lstick{$a_{1}: \ket{0}$}
        & \gate{H}
        & \ctrl{-5} &
        &           & \ctrl{-4}
        & \ctrl{-3} &
        &           & \ctrl{-2}
        & \gate{H}  & \meter{}
        & \cw
    \end{quantikz}
\end{align}
Similarly introducing \begin{quantikz}[{column sep=2pt}]&\gate{{\rm QED}_1}&\end{quantikz} with the stabilizers $S_{X1}$ and $S_{Z1}$, we define the FT QED gadget used in the main text by
\begin{align}
\label{eq:circ_QED}
    \begin{quantikz}[style={row sep=5pt, column sep=4pt}]
        \lstick{}          & \gate{\rm QED} & \qw
    \end{quantikz}
    =
    \begin{quantikz}[style={row sep=5pt, column sep=4pt}]
        \lstick{}          & \gate{{\rm QED}_0} & \gate{{\rm QED}_1} & \qw
    \end{quantikz}
\end{align}

\end{itemize}

\subsection{Non-fault-tolerant \texorpdfstring{$Z$}{Z} rotation}
\label{app:Z_rotation}

We consider the non-fault-tolerant $\overline{R}(\theta)$ gate
\begin{align}
\label{eq:R_Z^D}
    \begin{quantikz}[style={column sep=4pt, row sep=5pt}]
        \lstick{} & \gate[style={fill=blue!10}]{\overline{R}_{Z}^{\rm D}(\theta)} & \qw
    \end{quantikz}
    ~=~
    \begin{quantikz}[style={column sep=4pt, row sep=5pt}]
        \lstick{$q_{0}$} & \qw       & \gate[2]{R_{ZZ}(\theta)} & \qw       & \qw \\
        \lstick{$q_{1}$} & \targ{}   &                         & \targ{}   & \qw \\
        \lstick{$q_{4}$} & \ctrl{-1} & \qw                     & \ctrl{-1} & \qw \\
    \end{quantikz}
\end{align}
with $R_{ZZ}(\theta)=\e^{-\im\frac{\theta}{2}ZZ}$, and analyze how the errors that occur after each two-qubit gate propagate through other gates in the circuit with probability $p$.
We assume any Pauli error up to weight two occurs after each gate, indicated by $(i)$, $(ii)$, and $(iii)$,
\begin{align}
    \begin{quantikz}[style={column sep=8pt, row sep=6pt}, slice style=black]
         \lstick{$q_0$}
         & \slice{$(i)$}
         & \gate[2]{R_{ZZ}(\theta)} \slice{$(ii)$}
         & \slice{$(iii)$}
         &
         \\
         \lstick{$q_1$}
         & \targ{}
         &
         & \targ{}
         &
         \\
         \lstick{$q_4$}
         & \ctrl{-1}
         &
         & \ctrl{-1}
         &
    \end{quantikz}
\end{align}
The errors ($E$) that occur at the slices and resultant errors ($O$) after propagating them through are summarized in Tab.~\ref{tab:errors}.
Three remarks are in order: an $X_1$ error on the way in, or from the first $CX$, will flip the logical gate and cannot be distinguished from the same error resulting from a failure in a later gate.
$\overline{Z}$ errors can occur with probability ${\cal O}(p)$, but only as the result of a $ZZ$ error after gate $(ii)$.
Weight-two errors can occur with probability ${\cal O}(p)$.

Following the partially fault-tolerant design principle, we append the gate with the QED gadget
\eqref{eq:circ_QED},
\begin{align}
\label{eq:R_Z^DQED}
    \begin{quantikz}[style={column sep=4pt, row sep=5pt}]
        &\gate[style={fill=blue!10}]{\overline{R}_{Z}^{\mathrm{D}}(\theta)} & \gate{\rm QED} & \qw
    \end{quantikz}
\end{align}
Syndromes obtained by measuring the stabilizers $S_{X0}$, ($S_{X1}$), $S_{Z0}$, and $S_{Z1}$ in the QED gadget are given Tab.~\ref{tab:errors}.
The syndromes associated with $S_{X1}$ are shown with parentheses, without which one can detect all weight-two errors.

\begin{table}[!htbp]
\centering
\def\arraystretch{1.5}
\begin{tabular}{cccc|cccc|cccc}
    \hline
    \multicolumn{4}{c|}{$(i)$}
    & \multicolumn{4}{c|}{$(ii)$}
    & \multicolumn{4}{c}{$(iii)$}
    \\
    \quad $E$ \enspace & \quad $O$ \enspace & Flip & \quad Syndrome \enspace
    & \quad $E$ \enspace & \quad $O$ \enspace & Flip & \quad Syndrome \enspace
    & \quad $E$ \enspace & \quad $O$ \enspace & Flip & \quad Syndrome \enspace
    \\\hline
      $X_1$    & $X_1$       & 1 & 0(0)11
    & $X_0$    & $X_0$       & 0 & 0(0)10
    & $X_1$    & $X_1$       & 0 & 0(0)11
    \\
      $Y_1$    & $Y_1Z_4$    & 1 & 1(0)11
    & $Y_0$    & $Y_0$       & 0 & 1(0)10
    & $Y_1$    & $Y_1$       & 0 & 1(1)11
    \\
      $Z_1$    & $Z_1Z_4$    & 0 & 1(0)00
    & $Z_0$    & $Z_0$       & 0 & 1(0)00
    & $Z_1$    & $Z_1$       & 0 & 1(1)00
    \\
      $X_4$    & $X_1X_4$    & 0 & 0(0)10
    & $X_1$    & $X_1$       & 0 & 0(0)11
    & $X_4$    & $X_4$       & 0 & 0(0)01
    \\
      $X_1X_4$ & $X_4$       & 1 & 0(0)01
    & $X_0X_1$ & $X_0X_4$    & 0 & 0(0)11
    & $X_1X_4$ & $X_1X_4$    & 0 & 0(0)10
    \\
      $Y_1X_4$ & $Z_1Y_4$    & 1 & 1(0)01
    & $Y_0X_1$ & $Y_0X_1$    & 0 & 1(0)01
    & $Y_1X_4$ & $Y_1X_4$    & 0 & 1(1)10
    \\
      $Z_1X_4$ & $Y_1Y_4$    & 0 & 1(0)10
    & $Z_0X_1$ & $Z_0X_1$    & 0 & 1(0)11
    & $Z_1X_4$ & $Z_1X_4$    & 0 & 1(1)01
    \\
      $Y_4$    & $X_1Y_4$    & 0 & 0(1)10
    & $Y_1$    & $Y_1Z_4$    & 0 & 1(0)11
    & $Y_4$    & $Y_4$       & 0 & 0(1)01
    \\
      $X_1Y_4$ & $Y_4$       & 1 & 0(1)01
    & $X_0Y_1$ & $X_0Y_1Z_4$ & 0 & 1(0)01
    & $X_1Y_4$ & $X_1Y_4$    & 0 & 0(1)10
    \\
      $Y_1Y_4$ & $Z_1X_4$    & 1 & 1(1)01
    & $Y_0Y_1$ & $Y_0Y_1Z_4$ & 0 & 0(0)01
    & $Y_1Y_4$ & $Y_1Y_4$    & 0 & 1(0)10
    \\
      $Z_1Y_4$ & $Y_1X_4$    & 0 & 1(1)10
    & $Z_0Y_1$ & $Z_0Y_1Z_4$ & 0 & 0(0)11
    & $Z_1Y_4$ & $Z_1Y_4$    & 0 & 1(0)01
    \\
      \bm{$Z_4$}    & \bm{$Z_4$}       & \bf{0} & \bf{0(1)00}
    & $Z_1$    & $Z_1Z_4$    & 0 & 1(0)00
    & \bm{$Z_4$}    & \bm{$Z_4$}       & 0 & \bf{0(1)00}
    \\
      $X_1Z_4$ & $X_1Z_4$    & 1 & 0(1)11
    & $X_0Z_1$ & $X_0Z_1Z_4$ & 0 & 1(0)10
    & $X_1Z_4$ & $X_1Z_4$    & 0 & 0(1)11
    \\
      $Y_1Z_4$ & $Y_1$       & 1 & 1(1)11
    & $Y_0Z_1$ & $Y_0Z_1Z_4$ & 0 & 0(0)10
    & $Y_1Z_4$ & $Y_1Z_4$    & 0 & 1(0)11
    \\
      $Z_1Z_4$ & $Z_1$       & 0 & 1(1)00
    & \bm{$Z_0Z_1$} & \bm{$Z_0Z_1Z_4$} & 0 & \bf{0(0)00}
    & $Z_1Z_4$ & $Z_1Z_4$    & 0 & 1(0)00
    \\\hline
\end{tabular}
\caption{\label{tab:errors}
    Errors $E$ that occur at slices $(i)$-$(iii)$  in the circuit, along with output Paulis $O$. Flip indicates whether the logical rotation changes direction. Syndromes are for stabilizers $S_{X0}$, ($S_{X1}$), $S_{Z0}$, and $S_{Z1}$ in that order.
    The errors that incur the $000$ syndrome by measuring $(S_{X0}, S_{Z0}, S_{Z1})$ are shown in bold.
}
\end{table}

\subsection{Preparation of \texorpdfstring{$\ket{\overline{\theta}}$}{the logical theta state}}

We show quantum circuits to encode the state $\ket{\overline{\theta}} := \overline{R}_{Z}(\theta)\ket{\overline{+}}$.

\begin{itemize}

\item State preparation without error detection

The non-FT $\ket{\overline{\theta}}$ preparation circuit is expressed as
\begin{align}
\label{eq:nonFT_resource_prep}
    \ket{\overline{\theta}^{\rm noQED}}
    ~=~
    \begin{quantikz}[style={column sep=4pt, row sep=5pt}]
        \lstick{$\ket{0}$} &  \gate[1]{\text{$H$}} & \ctrl{1}  & \ctrl{3} & \qw       & \qw       & \qw                      & \qw        & \gate{H} & \qw \\
        \lstick{$\ket{0}$} &  \qw                  & \targ{}   & \qw      & \qw       & \targ{}   & \qw                      & \qw        & \gate{H} & \qw \\
        \lstick{$\ket{0}$} &  \qw                  & \qw       & \qw      & \targ{}   & \qw       & \qw                      & \targ{}    & \gate{H} & \qw \\
        \lstick{$\ket{0}$} &  \qw                  & \targ{}   & \targ{}  & \qw       & \qw       & \gate[2]{R_{XX}(\theta)} & \ctrl{-1}  & \gate{H} & \qw \\
        \lstick{$\ket{0}$} &  \gate[1]{\text{$H$}} & \qw       & \ctrl{1} & \ctrl{-2} & \ctrl{-3} &                          & \qw        & \gate{H} & \qw \\
        \lstick{$\ket{0}$} &  \qw                  & \qw       & \targ{}  & \targ{}   & \qw       & \qw                      & \qw        & \gate{H} & \qw \\
        \lstick{$\ket{0}$} &  \gate[1]{\text{$H$}} & \ctrl{-3} & \qw      & \ctrl{-1} & \qw       & \qw                      & \qw        & \gate{H} & \qw \\
    \end{quantikz}
\end{align}

\item State preparation with error detection

We use $\direct_Z(\theta)$ and error detection~\eqref{eq:R_Z^DQED} to prepare $\ket{\overline{\theta}}$,
\begin{align}
\label{eq:partialFT_resource_prep}
    \ket{\overline{\theta}^{\rm QED}}
    ~=~
    \begin{quantikz}[style={column sep=4pt, row sep=5pt}]
        \lstick{$\ket{\overline{+}^{\mathrm{FT}}}$} &\gate[style={fill=blue!10}]{\overline{R}_{Z}^{\mathrm{D}}(\theta)} & \gate{\rm QED} & \qw
    \end{quantikz}
\end{align}

{
\item Recursive gate teleportation (RGT)

The partially FT RGT used in our experiments applies QEDs on the preparation of the ancilla state $\ket{\overline{\theta}}$.
Errors in \eqref{eq:partialFT_resource_prep} can be detected either by the QED in the FT $\ket{\overline{+}}$ preparation~\eqref{eq:circ_prep_FT}, or the QEDs~\eqref{eq:circ_QED} trailing $\overline{R}_{Z}^{\mathrm{D}}(\theta)$. This is repeated until success with the maximum number of repetitions $S_{\rm max}$.
Having successfully prepared the ancilla state, we perform gate teleportation with
\begin{align}
\label{eq:R_Z^P_alg}
    \begin{quantikz}[style={column sep=4pt, row sep=7pt}]
        \lstick{$\ket{\psi}$}
        & \ctrl{1}
        & \qw
        & \rstick{$\overline{R}_Z(\theta)\ket{\psi}$\text{ or }$\overline{R}_Z(-\theta)\ket{\psi}$}\\
        \lstick{$\ket{\overline{\theta}}$}
        & \gate{\overline{X}}
        & \meterD{\overline{Z}}
        & \rstick{$m=\text{``0'' or ``1''}$}\cw \\
    \end{quantikz}
\end{align}
If it does not apply the desired operation on the system qubit, we start over from the ancilla state preparation with the doubled angle. This is repeated until success with the maximum number of repetitions $R_{\rm max}$.
These classically conditional operations are all executed at runtime as described in Algorithm~\ref{alg:RGT}.
}

\begin{algorithm}
{
\caption{RGT}
\SetKwInOut{Input}{input}
\SetKwInOut{Output}{output}
\SetKwInOut{Registers}{registers}
\SetKw{Run}{Run}
\label{alg:RGT}
    \Input{
        A rotation angle $\theta\in\mathbb{R}$,
        the maximum number of recursions $r_{\rm rus}\in\mathbb{N}$ for $\ket{\overline{\theta}}$ preparation,
        the maximum number of recursions $S_{\rm max}\in\mathbb{N}$ for gate teleportation.
    }
        \Registers{
        Input state $\ket{\psi}$ on system register $\mathtt{QREG_{\rm sys}}$,
        ancilla register $\mathtt{QREG_{\rm anc}}$.
    }
    $R\gets0$\;
    \While{$r<r_\mathrm{rus}$}{
        Initialize $\mathtt{QREG_{\rm anc}}$ to $\ket{\overline{+}^{\mathrm{FT}}}$\;
        $S\gets0$\;
        \While{$S<S_\mathrm{max}$}{
            \Run{the circuit~\eqref{eq:partialFT_resource_prep} on $\mathtt{QREG_{\rm anc}}$}\;
            \If{no error is detected in \eqref{eq:partialFT_resource_prep}}{
                \Run{the circuit~\eqref{eq:R_Z^P_alg} on $(\mathtt{QREG_{\rm sys}}, \mathtt{QREG_{\rm anc}}, \mathtt{CREG})$} \tcp*{$\ket{\bar{\theta}}$ is successfully prepared on $\mathtt{QREG_{\rm anc}}$}
                $m\gets$ measurement outcome of $\mathtt{QREG_{\rm anc}}$\;
                \eIf{$m=0$}{
                    \Output{$\mathtt{QREG_{\rm sys}}$ \tcp*{$R_Z(\theta)\ket{\psi}$ is successfully prepared on $\mathtt{QREG_{\rm sys}}$}}
                }{
                    $\theta\gets 2\theta$ \tcp*{$R_Z(-\theta)\ket{\psi}$ is prepared on $\mathtt{QREG_{\rm sys}}$}
                    $r\gets r+1$\;
                    Proceed to the next outer ($r$) loop\;
                }
            }
            $S\gets S+1$\;
        }
        Abort the circuit execution \tcp*{Discard the shot}
        \Output{failure}
    }
    Abort the circuit execution \tcp*{Discard the shot}
    \Output{failure}
}
\end{algorithm}

\end{itemize}

\subsection{Encoded QPE circuits}
\label{app:encoding}

We provide three encoding strategies of the QPE circuit~\eqref{eq:qpe_circ}, \textsf{PFT}, \textsf{Exp}, and \textsf{Dir}.
We call the setup employed in Sec.~\ref{sec:encoding} \textsf{PFT}, while \textsf{Exp} and \textsf{Dir} are the strategies used in the experiment presented in App.~\ref{app:non_ft}. {See Tab.~\ref{tab:resource_allocation} for the resource allocation of each logical component.}

\begin{table}[!htbp]
{
\centering
\def\arraystretch{1.5}
\begin{tabular}{c||c|c|c}
\hline
  & logical qubit count
  & physical qubit count
  & physical 2Q gate count
\\\hline\hline
  $\begin{quantikz}[style={column sep=2pt, row sep=2pt}]
    \lstick{} & \gate[style={fill=green!15}]{\rgt_{Z}(\theta)} & \qw
  \end{quantikz}$
  & 2 & 16 & 34 (183) \\\hline
  $\begin{quantikz}[style={column sep=2pt, row sep=2pt}]
    \lstick{} & \gate[style={fill=yellow!40}]{\rgt_Z(\theta)} & \qw
  \end{quantikz}$
  & 2 & 14 & 16 (48)  \\\hline
  $\begin{quantikz}[style={column sep=2pt, row sep=2pt}]
    \lstick{} & \gate[style={fill=blue!10}]{\direct_Z(\theta)} & \qw
  \end{quantikz}$
  & 1 & 7 & 3  \\\hline
  $\begin{quantikz}[style={column sep=2pt, row sep=2pt}]
    \lstick{} & \gate{{\rm QEC}_X}
    & \qw
  \end{quantikz}$  & 2 & 15 & 18 (29) \\\hline
\end{tabular}
\caption{\label{tab:resource_allocation}
    Qubit and gate counts for each logical component.
}
}
\end{table}

\begin{itemize}

\item \textsf{PFT}:
The encoded circuit is given by
\begin{align}
\label{eq:circ_QPE_PFT}
    \begin{quantikz}[style={column sep=4pt, row sep=8pt}]
        \lstick{$\ket{\overline{+}^{\rm FT}}$}
        & \ctrl{1}
        & \gate{\overline{R}_{Z}(\beta)}
        & \meterD{\overline{X}}
        \\
        \lstick{$\ket{\overline{\Phi}(\alpha_{0}, \alpha_{1})}$}
        & \gate{\overline{U}^{k}}
        &
    \end{quantikz}
\end{align}
where the Clifford gate $\overline{R}_{Z}(\beta)$ for $\beta\in\{0,\pi/2\}$ is implemented transversally.
The input state $\ket{\overline{\Phi}}$ is encoded as
\begin{align}
    \begin{quantikz}[style={column sep=4pt, row sep=8pt}]
        \lstick{$\ket{\overline{\Phi}(\alpha_{0}, \alpha_{1})}$}
    \end{quantikz}
    ~=~
    \begin{quantikz}[style={column sep=4pt, row sep=8pt}]
        \lstick{$\ket{\overline{0}^{\mathrm{FT}}}$}
        & \gate{\overline{X}}
        & \gate{\overline{H}}
        & \gate[style={fill=blue!10}]{\overline{R}_{Z}^{\rm D}(\alpha_{1})}
        & \gate{\overline{H}}
        & \gate{\overline{S}}
        & \gate[style={fill=blue!10}]{\overline{R}_{Z}^{\rm D}(\alpha_{2})}
        & \gate{\rm QED}
        &
    \end{quantikz}
\end{align}
The rotation gates in the phase-shift gate and $\ket{\overline{\Phi}}$ preparation are implemented by $\overline{R}_{Z}^{\rm D}$~\eqref{eq:R_Z^D} followed by a QED gadget~\eqref{eq:circ_QED}.
The logical controlled-$U^k$ gate, represented by a controlled-$\overline{U}^k$ in \eqref{eq:circ_QPE_PFT}, is given by repeating the following gate $k$ times,
\begin{align}
\label{eq:circ_ctrl_U_PFT}
    &\begin{quantikz}[style={column sep=4pt, row sep=7pt}]
        & \ctrl{1}            &
        \\
        & \gate{\overline{U}} &
        \\
    \end{quantikz}
    ~
    =
    \nonumber\\
    ~
    &\begin{quantikz}[style={column sep=2pt, row sep=7pt}]
        &
        & \ctrl{1}
        &
        & \ctrl{1}
        & \gate{\mathrm{QEC}_X}
        &
        &
        & \ctrl{1}
        &
        & \ctrl{1}
        &
        & \gate{\mathrm{QEC}_{X,Z}}
        &
        \\
        & \gate[style={fill=green!15}]{\rgtQED_Z(h_{1}t)}
        & \gate{\overline{X}}
        & \gate[style={fill=green!15}]{\rgtQED_Z(-h_{1}t)}
        & \gate{\overline{X}}
        & \gate{\mathrm{QEC}_X}
        & \gate{\overline{H}}
        & \gate[style={fill=green!15}]{\rgtQED_Z(h_{2}t)}
        & \gate{\overline{X}}
        & \gate[style={fill=green!15}]{\rgtQED_Z(-h_{2}t)}
        & \gate{\overline{X}}
        & \gate{\overline{H}}
        & \gate{\mathrm{QEC}_{X,Z}}
        &
    \end{quantikz}
\end{align}
where the rotation gates are implemented partially fault-tolerantly:
\begin{align}
\label{eq:R_Z^P}
    \begin{quantikz}[style={column sep=4pt, row sep=5pt}]
        \lstick{} & \gate[style={fill=green!15}]{\rgtQED_Z(\theta)} & \qw
    \end{quantikz}
    ~=~
    \begin{quantikz}[style={column sep=4pt, row sep=5pt}]
        \lstick{}                          & \ctrl{1}            & \gate[style={fill=green!15}]{\rgtQED_Z(2\theta)} & \qw \\
        \lstick{$\ket{\overline{\theta}^{\rm QED}}$} & \gate{\overline{X}} & \meterD{\overline{Z}} \vcw{-1} \\
    \end{quantikz}
\end{align}

\item \textsf{Exp}: This strategy removes the circuit overhead of \texttt{PFT} by sacrificing partially FT designs.
More specifically, the number of Steane QEC gadgets is reduced~\eqref{eq:circ_ctrl_U_Exp}, and the RGT gadget consumes $\ket{\overline{\theta}^{\rm noQED}}$ that is prepared without QED~\eqref{eq:R_Z^M}.
The encoded circuit,
\begin{align}
    \begin{quantikz}[style={column sep=4pt, row sep=8pt}]
        \lstick{$\ket{\overline{+}^{\rm nonFT}}$}                            & \ctrl{1}                & \gate[style={fill=blue!10}]{\overline{R}_{Z}^{\rm D}(\beta-h_{2}kt)} & \meterD{\overline{X}}
        \\
        \lstick{$\ket{\overline{\Phi}(\alpha_{0}, \alpha_{1})}$} & \gate{\overline{U}^{k}} & \qw
    \end{quantikz}
\end{align}
with
\begin{align}
    \begin{quantikz}[style={column sep=4pt, row sep=8pt}]
        \lstick{$\ket{\overline{\Phi}(\alpha_{0}, \alpha_{1})}$}
    \end{quantikz}
    ~=~
    \begin{quantikz}[style={column sep=4pt, row sep=8pt}]
        \lstick{$\ket{\overline{0}}$}
        & \gate{\overline{X}}
        & \gate{\overline{H}}
        & \gate[style={fill=blue!10}]{\overline{R}_{Z}^{\rm D}(\alpha_{1})}
        & \gate{\overline{H}}
        & \gate{\overline{S}}
        & \gate[style={fill=blue!10}]{\overline{R}_{Z}^{\rm D}(\alpha_{2})}
        &
    \end{quantikz}
\end{align}
is similar to that of \textsf{PFT}~\eqref{eq:circ_QPE_PFT}, {but the QED gadget is removed}.
Furthermore, $\overline{{\rm ctrl-}U}$ is implemented by
\begin{align}
\label{eq:circ_ctrl_U_Exp}
\begin{split}
    &\begin{quantikz}[style={column sep=4pt, row sep=7pt}]
        & \ctrl{1} &
        \\
        & \gate{\overline{U}} &
    \end{quantikz}
    ~
    =
    ~
    \\
    &\begin{quantikz}[style={column sep=3pt, row sep=7pt}]
        &
        & \ctrl{1}
        &
        & \ctrl{1}
        & \gate{\mathrm{QEC}_X}
        &
        &
        & \ctrl{1}
        &
        & \ctrl{1}
        &
        & \gate{\mathrm{QEC}_{X,Z}}
        &
        \\
        & \gate[style={fill=yellow!40}]{\rgt_Z(h_{1}t)}
        & \gate{\overline{X}}
        & \gate[style={fill=yellow!40}]{\rgt_Z(-h_{1}t)}
        & \gate{\overline{X}}
        & \gate{\mathrm{QEC}_X}
        & \gate{\overline{H}}
        & \gate[style={fill=yellow!40}]{\rgt_Z(h_{2}t)}
        & \gate{\overline{X}}
        & \gate[style={fill=yellow!40}]{\rgt_Z(h_{2}t)}
        & \gate{\overline{X}}
        & \gate{\overline{H}}
        & \gate{\mathrm{QEC}_{X,Z}}
        &
    \end{quantikz}
\end{split}
\end{align}
where the rotation gates are given by the following RGT
\begin{align}
\label{eq:R_Z^M}
    \begin{quantikz}[style={column sep=4pt, row sep=5pt}]
        \lstick{} & \gate[style={fill=yellow!40}]{\rgt_Z(\theta)} & \qw
    \end{quantikz}
    ~=~
    \begin{quantikz}[style={column sep=4pt, row sep=5pt}]
        \lstick{}                          & \ctrl{1}            & \gate[style={fill=yellow!40}]{\rgt_Z(2\theta)} & \qw
        \\
        \lstick{$\ket{\overline{\theta}^{\rm noQED}}$} & \gate{\overline{X}} & \meterD{\overline{Z}} \vcw{-1} \\
    \end{quantikz}
\end{align}
Compared to the RGT~\eqref{eq:R_Z^P}, the state $\ket{\overline{\theta}}$ is prepared with the circuit~\eqref{eq:nonFT_resource_prep} to reduce the circuit overhead at the cost of allowing more uncorrectable errors.
Finally, the circuit~\eqref{eq:circ_ctrl_U_Exp} has fewer QEC gadgets than the one for \textsf{PFT}~\eqref{eq:circ_ctrl_U_PFT}.

{
\item \textsf{Dir}: The encoded QPE circuit for \textsf{Dir} implements the logical $Z$-rotation gates with the non-FT circuit~\eqref{eq:R_Z^D},
\begin{align}
\begin{split}
    &\begin{quantikz}[style={column sep=4pt, row sep=7pt}]
        & \ctrl{1} &
        \\
        & \gate{\overline{U}} &
    \end{quantikz}
    ~
    =
    ~
    \\
    &\begin{quantikz}[style={column sep=3pt, row sep=7pt}]
        &
        & \ctrl{1}
        &
        & \ctrl{1}
        & \gate{\mathrm{QEC}_X}
        &
        &
        & \ctrl{1}
        &
        & \ctrl{1}
        &
        & \gate{\mathrm{QEC}_{X,Z}}
        &
        \\
        & \gate[style={fill=blue!10}]{\overline{R}_{Z}^{\rm D}(h_{1}t)}
        & \gate{\overline{X}}
        & \gate[style={fill=blue!10}]{\overline{R}_{Z}^{\rm D}(-h_{1}t)}
        & \gate{\overline{X}}
        & \gate{\mathrm{QEC}_X}
        & \gate{\overline{H}}
        & \gate[style={fill=blue!10}]{\overline{R}_{Z}^{\rm D}(h_{2}t)}
        & \gate{\overline{X}}
        & \gate[style={fill=blue!10}]{\overline{R}_{Z}^{\rm D}(h_{2}t)}
        & \gate{\overline{X}}
        & \gate{\overline{H}}
        & \gate{\mathrm{QEC}_{X,Z}}
        &
    \end{quantikz}
\end{split}
\end{align}
}

\end{itemize}

\section{Additional experimental data}
\label{app:non_ft}

\begin{figure}[!htbp]
    {
    \centering
    \includegraphics[width=0.5\linewidth]{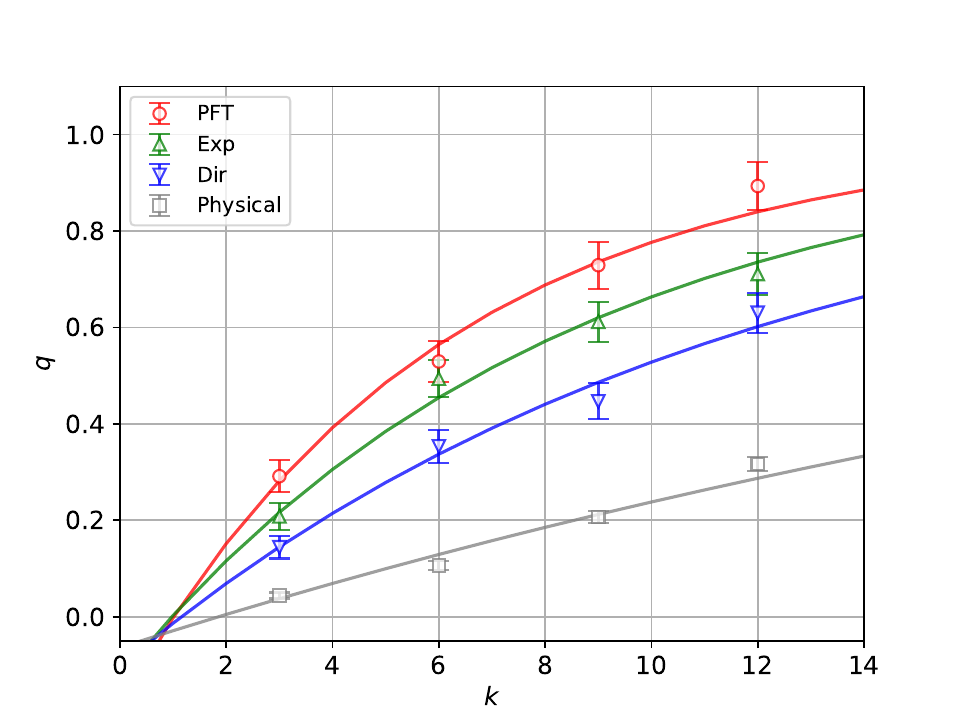}
    \caption{Decoherence parameter $q$ obtained by fitting the probability of zero outcome from the QPE circuits with $P^{(\mathrm{calc})}$ for the different implementations.}
    \label{fig:calibration_comparison}
    }
\end{figure}

We provide additional experimental data {to conduct a comparative study of \textsf{PFT}, \textsf{Exp}, \textsf{Dir} encoding strategies along with the unencoded setup (\textsf{Physical}) (see App.~\ref{app:encoding} for the setup detail).
In order to benchmark the performance of the QPE circuit under these setups, we quantify the circuit error (Sec.~\ref{ssec:calibration}) by running the calibration circuit discussed in Sec.~\ref{subsec:calibration}.
Figure~\ref{fig:calibration_comparison} shows the error parameter $q$ extracted by running the (un)encoded circuits.
We observe that the setup with less circuit overhead suffers from less hardware noise despite lacking fault tolerance.}

\section{Default noise parameters of H2 emulator}
\label{app:noise_params}

\begin{table}[!htbp]
    \centering
    \def\arraystretch{1.5}
    \begin{tabular}{l|c||c}
        \hline
        \multicolumn{2}{c||}{Noise model} & Parameter
        \\
        \hline\hline
        \multicolumn{2}{c||}{1-qubit-gate fault ($p_1$)} & $7.3\times10^{-5}$
        \\
        \hline
        \multicolumn{2}{c||}{two-qubit-gate fault ($p_2$)} & $1.29\times10^{-3}$
        \\
        \hline
        \multicolumn{2}{c||}{Initialization fault ($p_{\rm init}$)} & $4\times10^{-5}$
        \\
        \hline
        \multirow{2}{*}{Readout fault}& ``1''$\to$``0'' ($p_{\rm r, 1\to0}$) & $1.8\times10^{-3}$
        \\
        \cline{2-3}
        & ``0''$\to$``1'' ($p_{\rm r,0\to1}$) & $0.9\times10^{-3}$
        \\
        \hline
        \multirow{2}{*}{Memory noise rate} & coherent ($f$) & $4.3\times10^{-2}\text{rad}\cdot\text{s}^{-1}$
        \\
        \cline{2-3}
        & incoherent ($g$) & $2.8\times10^{-3}\text{s}^{-1}$
        \\
        \hline
    \end{tabular}
    \caption{\label{tab:default_noise_parameters}
    The default noise parameters used in our numerical simulations with the H2 emulator. The memory noise rates $f$ and $g$ are defined in~\eqref{eq:memory_noise}.
    }
\end{table}

Table~\ref{tab:default_noise_parameters} provides some of the default noise parameters used in the H2 emulator when the simulations are carried out.
The one and two-qubit gate faults are modeled by anisotropic Pauli channels.
The initialization of a qubit suffers from a bit-flip with the probability $p_{\rm init}$.
The readout fault flips a result from ``1'' to ``0'' with the probability $p_{\rm r, 1\to0}$, and from ``0'' to ``1'' with the probability $p_{\rm r,0\to1}$.
The memory noise on idling or moving qubits is modeled by the channel~\eqref{eq:memory_noise}.
See~\cite{H2emulator} for a complete list of the noise models and corresponding parameters.

\end{document}